\documentstyle[12pt]{article}

\newskip\humongous \humongous=0pt plus 1000pt minus 1000pt

\newif\ifdtup

\def\Re{\mathop{\rm Re}}

\def\abs#1{\left| #1\right|}
\def\pr#1{#1^\prime}

\def\beq{\begin{equation}}
\def\eeq{\end{equation}}

\def\beqn{\begin{eqnarray}}
\def\eeqn{\end{eqnarray}}
\relax

\def\dotx{\dotx{\dot\overline{x}}}

\relax

\catcode`\@=11

\@addtoreset{equation}{section}
\def\theequation{\thesection\arabic{equation}}

\def\@normalsize{\@setsize\normalsize{15pt}\xiipt\@xiipt
\abovedisplayskip 14pt plus3pt minus3pt%
\belowdisplayskip \abovedisplayskip
\abovedisplayshortskip \z@ plus3pt%
\belowdisplayshortskip 7pt plus3.5pt minus0pt}

\def\small{\@setsize\small{13.6pt}\xipt\@xipt
\abovedisplayskip 13pt plus3pt minus3pt%
\belowdisplayskip \abovedisplayskip
\abovedisplayshortskip \z@ plus3pt%
\belowdisplayshortskip 7pt plus3.5pt minus0pt
\def\@listi{\parsep 4.5pt plus 2pt minus 1pt
     \itemsep \parsep
     \topsep 9pt plus 3pt minus 3pt}}

\@twosidetrue

\relax

\catcode`@=12

\catcode`\@=11

\def\section{\@startsection{section}{1}{\z@}{3.5ex plus 1ex minus
   .2ex}{2.3ex plus .2ex}{\large\bf}}

\def\thesection{\arabic{section}.}

\def\appendix{\setcounter{section}{0}
 \def\thesection{APPENDIX \Alph{section}:}
 \def\theequation{\Alph{section}.\arabic{equation}}}

\def\ps@headings{\def\@oddfoot{}\def\@evenfoot{}
\def\@oddhead{\hbox{}\hfill
 \makebox[.5\textwidth]{\raggedright\ignorespaces --\thepage{}--
 \hfill {}}}  
\def\@evenhead{\@oddhead}
\def\subsectionmark##1{\markboth{##1}{}}
}

\ps@headings

\catcode`\@=12

\def\figcap{\section*{Figure Captions\markboth
 {FIGURECAPTIONS}{FIGURECAPTIONS}}\list
 {Fig. \arabic{enumi}:\hfill}{\settowidth\labelwidth{Fig. 999:}
 \leftmargin\labelwidth
 \advance\leftmargin\labelsep\usecounter{enumi}}}
 \relax
\def\tablecap{\section*{Table Captions\markboth
 {TABLECAPTIONS}{TABLECAPTIONS}}\list
 {Table \arabic{enumi}:\hfill}{\settowidth\labelwidth{Table 999:}
 \leftmargin\labelwidth
 \advance\leftmargin\labelsep\usecounter{enumi}}}
 \relax
\def\reflist{\section*{References\markboth
 {REFLIST}{REFLIST}}\list
 {[\arabic{enumi}]\hfill}{\settowidth\labelwidth{[999]}
 \leftmargin\labelwidth
 \advance\leftmargin\labelsep\usecounter{enumi}}}
 \relax

\catcode`\@=11

\def\ps@headings{\def\@oddfoot{}\def\@evenfoot{}
\def\@oddhead{\hbox{}\hfill
 \makebox[.5\textwidth]{\raggedright\ignorespaces --\thepage{}--
 \hfill {}}}    
\def\@evenhead{\@oddhead}
\def\subsectionmark##1{\markboth{##1}{}}
}

\ps@headings

\catcode`\@=12

\relax

\catcode`\@=11
\def\prm{\fam \z@}
\catcode`\@=12

\relax
\def\pl#1#2#3{{\it Phys. Lett. }{\bf #1}(19#2)#3}
\def\zp#1#2#3{{\it Z. Phys. }{\bf #1}(19#2)#3}
\def\prl#1#2#3{{\it Phys. Rev. Lett. }{\bf #1}(19#2)#3}

\def\prep#1#2#3{{\it Phys. Rep. }{\bf #1}(19#2)#3}
\def\pr#1#2#3{{\it Phys. Rev. }{\bf #1}(19#2)#3}
\def\np#1#2#3{{\it Nucl. Phys. }{\bf #1}(19#2)#3}

\def\ar#1#2#3{{\it Ann. Rev. Nucl. Part. Sc. }{\bf #1}(19#2)#3}

\relax

  \newcommand{\ccaption}[2]{
    \begin{center}
    \parbox{0.85\textwidth}{
      \caption[#1]{\small{\it{#2}}}
      }
    \end{center}
    }
\begin{document}
\newcommand\sss{\scriptscriptstyle}
\newcommand\mug{\mu_\gamma}
\newcommand\mue{\mu_e}
\newcommand\muf{\mu_{\sss F}}
\newcommand\mur{\mu_{\sss R}}
\newcommand\muo{\mu_0}
\newcommand\me{m_e}
\newcommand\as{\alpha_{\sss S}}
\newcommand\ep{\epsilon}
\newcommand\Th{\theta}
\newcommand\epb{\overline{\epsilon}}
\newcommand\aem{\alpha_{\rm em}}
\newcommand\refq[1]{$^{[#1]}$}
\newcommand\avr[1]{\left\langle #1 \right\rangle}
\newcommand\lambdamsb{\Lambda_5^{\rm \sss \overline{MS}}}
\newcommand\qqb{{q\overline{q}}}
\newcommand\qb{\overline{q}}
\newcommand\MSB{{\rm \overline{MS}}}
\newcommand\DIG{{\rm DIS}_\gamma}
\newcommand\CA{C_{\sss A}}
\newcommand\DA{D_{\sss A}}
\newcommand\CF{C_{\sss F}}
\newcommand\TF{T_{\sss F}}
\newcommand\Jetlist{\{J_l\}_{1,3}}
\newcommand\aoat{\sss a_{\sss 1} a_{\sss 2}}
\newcommand\SFfull{\{k_l\}_{1,6}}
\newcommand\SFfullbi{\{k_l\}_{1,6}^{[i]}}
\newcommand\SFTfull{\{k_l\}_{1,5}}
\newcommand\SFj{\{k_l\}_{3,6}}
\newcommand\FLfull{\{a_l\}_{1,6}}
\newcommand\FLfullbi{\{a_l\}_{1,6}^{[i]}}
\newcommand\FLTfull{\{a_l\}_{1,5}}
\newcommand\FLFj{\{a_l\}_{3,6}}
\newcommand\SFjbi{\{k_l\}_{3,6}^{[i]}}
\newcommand\FLjbi{\{a_l\}_{3,6}^{[i]}}
\newcommand\SFjexcl{\{k_l\}_{i,j}^{[np..]}}
\newcommand\SCollfull{\{k_l\}_{1,7}^{[ij]}}
\newcommand\SColl{\{k_l\}_{3,7}^{[ij]}}
\newcommand\FLCollfull{\{a_l\}_{1,7}^{[ij]}}
\newcommand\FLColl{\{a_l\}_{3,7}^{[ij]}}
\newcommand\STj{\{k_l\}_{3,5}}
\newcommand\FLTj{\{a_l\}_{3,5}}
\newcommand\Argfull{\FLfull;\SFfull}
\newcommand\ArgTfull{\FLTfull;\SFTfull}
\newcommand\KtoKF{(k_1,k_2\to\SFj\,;\FLFj)}
\newcommand\KtoKT{(k_1,k_2\to\STj\,;\FLTj)}
\newcommand\FLsum{\sum_{\{a_l\}}}
\newcommand\FLFsum{\sum_{\FLFj}}
\newcommand\FLFsumbi{\sum_{\FLjbi}}
\newcommand\FLTsum{\sum_{\FLTj}}
\newcommand\MF{{\cal M}^{(4)}}
\newcommand\MT{{\cal M}^{(3)}}
\newcommand\MTz{{\cal M}^{(3,0)}}
\newcommand\MTo{{\cal M}^{(3,1)}}
\newcommand\MTi{{\cal M}^{(3,i)}}
\newcommand\MTmn{{\cal M}^{(3,0)}_{mn}}
\newcommand\MFsj{\MF(k_1,k_2\to\SFj)}
\newcommand\MTsj{\MT(k_1,k_2\to\STj)}
\newcommand\MTisj{\MTi(k_1,k_2\to\STj)}
\newcommand\PHIFsj{\phi_4(k_1,k_2\to\SFj)}
\newcommand\PHITsj{\phi_3(k_1,k_2\to\STj)}
\newcommand\uoffct{\frac{1}{4!}}
\newcommand\uotfct{\frac{1}{3!}}
\newcommand\uoxic{\left(\frac{1}{\xi}\right)_c}
\newcommand\uoxiic{\left(\frac{1}{\xi_i}\right)_c}
\newcommand\uoxilc{\left(\frac{\log\xi}{\xi}\right)_c}
\newcommand\uoxiilc{\left(\frac{\log\xi_i}{\xi_i}\right)_c}
\newcommand\uoyim{\left(\frac{1}{1-y_i}\right)_+}
\newcommand\uoyimdi{\left(\frac{1}{1-y_i}\right)_{\delta_{\sss I}}}
\newcommand\uoyjmdo{\left(\frac{1}{1-y_j}\right)_{\delta_o}}
\newcommand\uoyip{\left(\frac{1}{1+y_i}\right)_+}
\newcommand\uoyipdi{\left(\frac{1}{1+y_i}\right)_{\delta_{\sss I}}}
\newcommand\uoyilm{\left(\frac{\log(1-y_i)}{1-y_i}\right)_+}
\newcommand\uoyilp{\left(\frac{\log(1+y_i)}{1+y_i}\right)_+}
\newcommand\uozm{\left(\frac{1}{1-z}\right)_+}
\newcommand\uozlm{\left(\frac{\log(1-z)}{1-z}\right)_+}
\newcommand\SVfact{\frac{(4\pi)^\ep}{\Gamma(1-\ep)}
                   \left(\frac{\mu^2}{Q^2}\right)^\ep}
\newcommand\gs{g_{\sss S}}
\newcommand\Icol{\{d_l\}}
\newcommand\An{{\cal A}^{(n)}}
\newcommand\Mn{{\cal M}^{(n)}}
\newcommand\Nn{{\cal N}^{(n)}}
\newcommand\Anu{{\cal A}^{(n-1)}}
\newcommand\Mnu{{\cal M}^{(n-1)}}
\newcommand\Sumae{\sum_{d_{e}}}
\newcommand\Sumaecl{\sum_{d_{e},\Icol}}
\newcommand\Sumaeae{\sum_{d_{e},d_{e}^{\prime}}}
\newcommand\Sumhe{\sum_{h_{e}}}
\newcommand\Sumhep{\sum_{h_{e}^\prime}}
\newcommand\Sumhehe{\sum_{h_{e},h_{e}^{\prime}}}
\newcommand\Pgghhh{S_{gg}^{h_e h_i h_j}}
\newcommand\Pggplus{S_{gg}^{+ h_i h_j}}
\newcommand\Pggminus{S_{gg}^{- h_i h_j}}
\newcommand\Pgghphh{S_{gg}^{h_e^\prime h_i h_j}}
\newcommand\Pqgplus{S_{qg}^{+ h_i h_j}}
\newcommand\Pqgminus{S_{qg}^{- h_i h_j}}
\newcommand\Pgqplus{S_{gq}^{+ h_i h_j}}
\newcommand\Pgqminus{S_{gq}^{- h_i h_j}}
\newcommand\Pqqplus{S_{qq}^{+ h_i h_j}}
\newcommand\Pqqminus{S_{qq}^{- h_i h_j}}
\newcommand\Hnd{\{h_l\}}
\newcommand\LC{\stackrel{\sss i\parallel j}{\longrightarrow}}
\newcommand\LCu{\stackrel{\sss 1\parallel j}{\longrightarrow}}
\newcommand\Physvar{\{V_l\}}
\newcommand\Physcm{\{\bar{V}_l\}}
\newcommand\Partvar{\{v_l\}}
\newcommand\Partcm{\{\bar{v}_l\}}
\newcommand\Partvarcm{\{v_l(\bar{v})\}}
\newcommand\Partcmset{\{\bar{v}_l^{(i)}\}}
\newcommand\Partvarcmset{\{v_l(\bar{v}^{(i)})\}}
\def\A#1#2{\la#1#2\ra}
\def\B#1#2{[#1#2]}
\newcommand{\la}{\langle}
\newcommand{\ra}{\rangle}
\newcommand{\nn}{\nonumber}
\newcommand\Qb{\overline{Q}}
\renewcommand\topfraction{1}       
\renewcommand\bottomfraction{1}    
\renewcommand\textfraction{0}      
\setcounter{topnumber}{5}          
\setcounter{bottomnumber}{5}       
\setcounter{totalnumber}{5}        
\setcounter{dbltopnumber}{2}       
\newsavebox\tmpfig
\newcommand\settmpfig[1]{\sbox{\tmpfig}{\mbox{\ref{#1}}}}
%
\begin{titlepage}
\setcounter{footnote}{1}
\begingroup
\def\thefootnote{\fnsymbol{footnote}}
\vspace*{-2cm}
\begin{flushright}
ETH-TH/95-42 \\
SLAC-PUB-95-7073 \\
December 1995 \\
\end{flushright}
\vskip .5cm
\begin{center}
{\large\bf
Three-jet cross sections
to next-to-leading order
}
\vskip 1cm
{\large S. Frixione$^a$\,, Z. Kunszt} \\
\vskip 0.12cm
Theoretical Physics, ETH, Zurich, Switzerland \\
\vskip 0.35cm
{\large A. Signer$^{a}$} \\
\vskip 0.12cm
SLAC, PO Box 4349, Stanford, CA 94309
\vskip .4cm
\end{center}
\endgroup
\begin{abstract}
\noindent
One- and two-jet inclusive quantities in hadron collisions
have already been calculated to
next-to-leading order accuracy, using both the subtraction and
the cone method. Since the one-loop corrections have recently
been obtained for all five-parton amplitudes, three-jet inclusive
quantities can also be predicted to next-to-leading order.
The subtraction method presented in the literature is based on a systematic
use of boost-invariant kinematical variables, and therefore its
application to three-jet production is quite cumbersome.
In this paper we re-analyze
the subtraction method and point out the advantage of using angle
and energy variables. This leads to simpler results and it has complete
generality, extending its validity to $n$-jet production.
The formalism is also applicable to $n$-jet production in
$e^+e^-$ annihilation and in photon-hadron collisions.
All the analytical results necessary to construct an efficient numerical
program for next-to-leading order three-jet inclusive quantities in
hadroproduction are given explicitly. As new analytical result, we also
report the collinear limits of all the two-to-four processes.
\end{abstract}

\vskip0.7truecm
\noindent
{}~$^a$ Work supported by the National Swiss Foundation
\end{titlepage}
\setcounter{footnote}{0}
\section{Introduction}

The production of one or more jets is an important
phenomenon in large transverse momentum reactions at
hadron-hadron and photon-hadron colliders~[\ref{ppreview}].
Jet production has large rates and rich
final state structure, thus providing an excellent testing
ground for the predictions of perturbative QCD and a formidable
background to various new physics signals. It is therefore important to
have as significant theoretical predictions as possible
for all the measured distributions in jet production.

In the last decade the formalism of the next-to-leading
order perturbative QCD has successfully been
applied~[\ref{onejettheory}-\ref{twojettheory}]
to explain quantitatively the data on one-jet and two-jet inclusive
quantities obtained at SPS, TEVATRON and
HERA~[\ref{onejet5401800}-\ref{twojetexp}].
At the next-to-leading order the theoretical uncertainties can be
reduced below the experimental errors and impressive agreement
has been found between the data and the theoretical predictions
for all available inclusive one-jet or two-jet measurements.

We recall, however, that there are two embarrassing unresolved
discrepancies. First, the ramping run at TEVATRON could be used to test
Feynman scaling; the data show larger scaling violation effects than
predicted by the theory~[\ref{onejet5401800}]. Second, the tail of
the transverse energy distribution of single-inclusive jet production,
measured by the CDF collaboration~[\ref{onejetexp}], appears to be higher
than the next-to-leading order QCD prediction
(see, however, ref.~[\ref{hustonetal}]).

The success (and also the possible difficulties) of the next-to-leading
order description of one- and two-jet production, and the large number
of three-jet events collected by experimental
collaborations~[\ref{multijetexp}], motivate
the extension of the quantitative next-to-leading order analysis also
to the description of the three-jet data. The study of the ratio of three-
to two-jet production rates is expected to give a clean measurement of
$\as$ at hadron colliders.

The theoretical analysis of inclusive three-jet production is rather
complex. The five-parton amplitudes have to be calculated to next-to-leading
order. As a result of significant technical developments based
on the helicity method, supersymmetry and string theory, these calculations
have been performed and presently the theoretical inputs at the
amplitude level are fully available~[\ref{BDK5g}-\ref{ASthesis}].
The leading order six-parton amplitudes, which have been obtained
long time ago~[\ref{sixparton}-\ref{ManganoParkepr}], are also needed.
The relation of the physical cross section with the partonic
amplitudes, however, is rather involved, since the soft and collinear
singularities appearing in the next-to-leading order amplitude of the
$2\to 3$ parton scattering  and in the leading-order $2\to 4$
amplitudes are cancelled only when the cross section of inclusive,
infrared-safe quantities is considered.
In an efficient phenomenological numerical study, the soft and collinear
singularities appearing in the loop corrections and in the real
contributions have to be cancelled analytically. It has been pointed
out in refs.~[\ref{kssub}-\ref{GGcone}] that the necessary analytical
cancellation can always be achieved due to good universal properties
of the soft and collinear limiting behavior of partonic amplitudes.
In particular, two methods have been proposed: the so-called subtraction
method~[\ref{kssub},\ref{EKSgluon}] and the cone method~[\ref{GGcone}].
Both formalisms have successfully been applied to describe various
distributions measured in inclusive one-jet and two-jet production.

The aim of this paper is to present some further improvements
of the formalism of the subtraction method necessary for the
phenomenological study of three-jet inclusive quantities.
The subtraction method of  ref.~[\ref{kssub}] is based on a  systematic
use of boost-invariant kinematical variables when the reference to the
beam direction remains always explicit. As a result, the
soft integrals become more complicated than their inherent structure
would require and their treatment is somewhat cumbersome when
it is applied to three or more jet production.

We point out that the formalism of ref.~[\ref{kssub}] remains valid
also when energy and angle variables are used. In this case the soft
integrals become much simpler and the formalism is generally
valid for inclusive production of any number of jets.
The subtraction method with energy and angle variables has
already been applied to the description of other production
processes; for a formulation quite close to the one we use
in this paper, see ref.~[\ref{HVQetal}].

The cancellation of leading singularities does not completely
solve the numerical problems. The subtracted cross sections
may still have integrable square-root singularities, thus
resulting in an inefficient numerical evaluation.
In ref.~[\ref{kssub}] this difficulty has been overcome by
decomposing the cross section with partial fractioning
into terms of single-singular factors.
This method is not practical in the case of three-jet
production since it leads to very long algebraic formulae.
We shall show that such a decomposition is not actually necessary,
since the measurement function defining infrared-safe
inclusive jet quantities automatically splits the squared
amplitude into terms where only one (or at most two) singular
regions can contribute.

In this paper we summarize the results obtained for the next-to-leading
order five-parton amplitudes, we construct local subtraction
terms for the singular regions of the six-parton cross sections
and we calculate the contributions of the soft and collinear regions
analytically. With the latter result we have all the analytic formulae
at our disposal to construct an efficient numerical program for
the evaluation of three-jet production rates to next-to-leading order
accuracy.

Our paper is organized as follows.
In section 2 the organization of the calculation is presented:
our definitions and conventions are given,
the $\MSB$ collinear counterterms are explicitly worked out,
the adopted jet-finding algorithm is described
and the measurement functions are constructed.
In section 3 we remind the known results for the virtual
contribution.
In section 4 we deal with the real contribution, and we
show how to decompose it into single-singular factors.
We define the soft subtraction using angle and energy variables
and we evaluate the soft integrals analytically. We next turn
to the definition of the subtraction terms for the initial
and final state collinear singularities; all the needed
integrals are analytically calculated.
In section 5 we collect our results, and we show
that when we sum the contributions of the real term,
of the collinear counterterms and of the virtual term all
the singularities cancel. The numerical implementation of
our analytical results is also shortly discussed.
Section 6 contains our concluding remarks.
The soft and collinear integrals are collected in appendix A.
Finally, in appendix B we present a detailed description of the
collinear limits of the $n$-parton cross sections.
\section{The jet cross section}

\subsection{Introductory remarks}

Thanks to the factorization theorem~[\ref{CSS}], a generic differential
cross section in hadronic collisions can be written in the
following way
\beq
d\sigma^{(H_1 H_2)}(K_1,K_2)=\sum_{ab}\int dx_1 dx_2 f^{(H_1)}_a(x_1)
f^{(H_2)}_b(x_2)d\hat{\sigma}_{ab}(x_1 K_1,x_2 K_2)\,,
\label{factth}
\eeq
where $H_1$ and $H_2$ are the incoming hadrons, $K_1$ and $K_2$
their momentum, and the sum runs over all the parton flavours
which give a non-trivial contribution. The quantities
$d\hat{\sigma}_{ab}$ are the {\it subtracted} partonic cross sections,
in which the singularities due to collinear emission
of massless partons from the incoming partons have been cancelled
by some suitable counterterms.

We first express the subtracted cross sections in terms of
the unsubtracted ones, which can be directly calculated in perturbative
QCD. To this end, we have to write the collinear counterterms.
Due to universality, eq.~(\ref{factth}) applies also when
the incoming hadrons are formally substituted with partons.
In this case, we are also able to evaluate the partonic
densities, which at the next-to-leading order read
\beq
f^{(d)}_a(x)=\delta_{ad}\delta(1-x)-\frac{\as}{2\pi}
\left(\frac{1}{\epb}P_{ad}(x,0)-K_{ad}(x)\right)
+{\cal O}\left(\as^2\right),
\eeq
where $P_{ad}(x,0)$ are the Altarelli-Parisi kernels in four
dimensions (since we will usually work in $4-2\ep$ dimensions, the $0$
in the argument of $P_{ad}$ stands for $\ep=0$) and the functions
$K_{ad}$ depend upon the subtraction scheme in which the calculation
is carried out. For ${\rm \overline{MS}}$, $K_{ad}\equiv 0$. Writing
the perturbative expansion of the unsubtracted and subtracted partonic
cross sections at next-to-leading order as
\beq
d\sigma_{ab}=d\sigma_{ab}^{(0)}+d\sigma_{ab}^{(1)}\,,\;\;\;\;
d\hat{\sigma}_{ab}=d\hat{\sigma}_{ab}^{(0)}+d\hat{\sigma}_{ab}^{(1)}\,,
\label{decomposition}
\eeq
where the superscript 0 (1) denotes the leading (next-to-leading)
order contribution, we have
\beqn
d\hat{\sigma}_{ab}^{(0)}(k_1,k_2)&=&d\sigma_{ab}^{(0)}(k_1,k_2)
\\*
d\hat{\sigma}_{ab}^{(1)}(k_1,k_2)&=&d\sigma_{ab}^{(1)}(k_1,k_2)
+\frac{\as}{2\pi}\sum_d\int dx\left(\frac{1}{\epb}P_{da}(x,0)
-K_{da}(x)\right)d\sigma_{db}^{(0)}(xk_1,k_2)
\nonumber \\*&&
+\frac{\as}{2\pi}\sum_d\int dx\left(\frac{1}{\epb}P_{db}(x,0)
-K_{db}(x)\right)d\sigma_{ad}^{(0)}(k_1,xk_2)\,.
\label{counterterms}
\eeqn
The second and the third term in the RHS of eq.~(\ref{counterterms})
are the collinear counterterms we were looking for. Notice that in this
equation the Born terms $d\sigma^{(0)}$ are evaluated in
$4-2\ep$ dimensions.

For three-jet production, the leading-order cross section
can get contributions only from the two-to-three partonic
subprocesses. We write this contribution in the following way
\beqn
d\sigma_{\aoat}^{(0)}(k_1,k_2;\Jetlist)&=&\uotfct
\FLTsum\MTz(\ArgTfull)
\nonumber \\*&\times&
{\cal S}_3(\STj;\Jetlist)d\PHITsj\,.
\label{bornjetdef}
\eeqn
Here we denoted with
\beq
\Jetlist\,=\,\{J_1,\,J_2,\,J_3\}
\eeq
the set of the four-momenta of the jets. In the following, we will
almost always suppress the indication of the $\Jetlist$ dependence.
We have indicated
with $k_i$ and $a_i$ respectively the momentum and the
flavour of the parton number $i$ involved in the process; by
definition, partons $1$ and $2$ are the incoming ones.
In the sum $\FLTsum$ every $a_l$, with $3\leq l\leq 5$, takes the values
$g$, $u$, $\bar{u}$, and so on. To avoid overcounting in physical
predictions, we inserted the factor $1/3!$. To shorten as much as
possible the notation, we have collectively indicated the momenta as
\beq
\{k_l\}_{i,j}\,\equiv\,\{k_l\,|\,i\leq l\leq j\,\}\,.
\eeq
It will also turn useful to define
\beq
\SFjexcl\equiv \{k_l\,|\,i\leq l\leq j,\,\,l\neq n,\,l\neq p,\,..\}\,.
\eeq
The same notation will be used for flavours. The quantity
${\cal S}_3$ is the so-called measurement function, which defines
the infrared-safe jet observables in terms of the momenta of the
(unobservable) partons; we will describe it in more details in the
following. $\MTz$ is the two-to-three {\it leading-order} transition
amplitude squared, summed over final state and averaged over
initial state color and spin degrees of freedom, and multiplied
by the flux factor
\beq
\MTz=\frac{1}{2k_1\cdot k_2}\,\frac{1}{\omega(a_1)\omega(a_2)}
\sum_{\stackrel{color}{spin}}\abs{{\cal A}^{(tree)}(2\to 3)}^2\,,
\label{bornampdef}
\eeq
where $\omega(a)$ is the number of color and spin degrees
of freedom for the flavour $a$. We remember that, in $4-2\ep$
dimensions,
\beq
\omega(q)=2N_c\,,\;\;\;\;
\omega(g)=2(1-\ep)\DA\,,
\eeq
where $\DA=N_c^2-1$ is the dimension of the adjoint representation
of the color group $SU(N_c)$. Since the incoming partons can
play a very special r\^ole, we will also write the functional
dependence of $\MTz$ in the following way
\beq
\MTz(a_1,a_2,\FLTj;\,k_1,k_2,\STj)\,.
\eeq
The transition amplitude for processes in which only gluons and quarks
are involved is usually evaluated in the unphysical configuration
in which all the particles are outgoing. The amplitude
${\cal A}^{(tree)}(2\to 3)$ of eq.~(\ref{bornampdef}) can be obtained
from the amplitude
\beq
{\cal A}^{(tree)}(0\to 5)={\cal A}^{(tree)}
(\bar{a}_1,\bar{a}_2,\FLTj;\,-k_1,-k_2,\STj)
\label{crosssymm}
\eeq
simply by crossing (for details on crossing, see appendix B);
notice that eq.~(\ref{crosssymm}) is crossing invariant.
Finally, in eq.~(\ref{bornjetdef}) we denoted with $d\phi_3$ the full
(i.e., with the $\delta$ that enforces the conservation of four-momentum)
three-body phase space.

Coming to the next-to-leading order contribution, both the
two-to-three and two-to-four partonic subprocesses have to
be considered. As customary in perturbative QCD, we denote
as virtual the contribution of the former, and as real the
contribution of the latter:
\beq
d\sigma_{\aoat}^{(1)}=d\sigma_{\aoat}^{(v)}
+d\sigma_{\aoat}^{(r)}\,,
\label{realplusvirt}
\eeq
with
\beqn
d\sigma_{\aoat}^{(v)}(k_1,k_2;\Jetlist)&=&\uotfct
\FLTsum\MTo(\ArgTfull)
\nonumber \\*&\times&
{\cal S}_3(\STj;\Jetlist)d\PHITsj\,,
\label{virtjetdef}
\\
d\sigma_{\aoat}^{(r)}(k_1,k_2;\Jetlist)&=&\uoffct
\FLFsum\MF(\Argfull)
\nonumber \\*&\times&
{\cal S}_4(\SFj;\Jetlist)d\PHIFsj\,,
\label{realjetdef}
\eeqn
where ${\cal S}_4$ is the measurement function, analogous to
${\cal S}_3$, for four partons in the final state. $\MTo$ is due
to the loop contribution to the two-to-three subprocesses
\beqn
\MTo&=&\frac{1}{2k_1\cdot k_2}\,\frac{1}{\omega(a_1)\omega(a_2)}
\nonumber \\*&\times&
\sum_{\stackrel{color}{spin}}\Bigg[{\cal A}^{(tree)}(2\to 3)
\left({\cal A}^{(loop)}(2\to 3)\right)^*
+\left({\cal A}^{(tree)}(2\to 3)\right)^*
{\cal A}^{(loop)}(2\to 3)\Bigg],
\nonumber \\*&&
\label{virtampdef}
\eeqn
while $\MF$ is defined in terms of the two-to-four transition amplitude
\beq
\MF=\frac{1}{2k_1\cdot k_2}\,\frac{1}{\omega(a_1)\omega(a_2)}
\sum_{\stackrel{color}{spin}}\abs{{\cal A}^{(tree)}(2\to 4)}^2\,.
\label{realampdef}
\eeq
Notice that in eq.~(\ref{realjetdef}) the dependence upon the momentum
and flavour of the additional parton was inserted. Also, the measurement
function, as well as the combinatorial factor $1/4!$, had to be modified
with respect to eq.~(\ref{bornjetdef}) and eq.~(\ref{virtjetdef}).
The full four-body phase space was denoted with $d\phi_4$.

We can now go back to eq.~(\ref{counterterms}), to write explicitly
the collinear counterterms for the three-jet production. Using
eq.~(\ref{bornjetdef}) we get
\beqn
d\sigma_{\aoat}^{(cnt+)}&=&\uotfct\frac{\as}{2\pi}\sum_d\int dx
\left(\frac{1}{\epb}P_{da_1}(x,0)-K_{da_1}(x)\right)
\nonumber \\*&\times&
\FLTsum\MTz(d,a_2,\FLTj;xk_1,k_2,\STj)
{\cal S}_3 d\phi_3(xk_1,k_2\to\STj)\,,
\nonumber \\*&&
\label{cnt1}
\\*
d\sigma_{\aoat}^{(cnt-)}&=&\uotfct\frac{\as}{2\pi}\sum_d\int dx
\left(\frac{1}{\epb}P_{da_2}(x,0)-K_{da_2}(x)\right)
\nonumber \\*&\times&
\FLTsum\MTz(a_1,d,\FLTj;k_1,xk_2,\STj)
{\cal S}_3 d\phi_3(k_1,xk_2\to\STj)\,.
\nonumber \\*&&
\label{cnt2}
\eeqn
By construction, these quantities, when added to the unsubtracted
three-jet partonic cross section, must cancel the collinear
singularities coming from initial state emission.

\subsection{The jet-finding algorithm}

We now turn to the problem of the definition of the measurement
functions ${\cal S}_3$ and ${\cal S}_4$. As a preliminary, we need some
prescription defining the way in which unobservable partons
are eventually merged into physical jets. To define a jet in terms
of partons, it is customary to distinguish two separate steps:
the clustering algorithm, which decides whether a given set of partons
is mergeable into a jet, and the merging procedure, which defines
the jet momentum as a function of the parton momenta.
The clustering algorithm we choose
to use was introduced by Ellis and Soper~[\ref{ESalg}]. It is
a $k_{\sss T}$ algorithm specifically designed for hadron-hadron
collisions. It is formulated in terms of the transverse momenta
$k_{i{\sss T}}$ and of the lego plot distances $R_{ij}$ of
the final state partons
\beqn
d_i&=&k_{i{\sss T}}^2\,,
\\
R_{ij}&=&(\eta_i-\eta_j)^2+(\varphi_i-\varphi_j)^2\,,
\\
d_{ij}&=&min(k_{i{\sss T}}^2,k_{j{\sss T}}^2)\frac{R_{ij}}{D^2}\,,
\eeqn
where the constant $D$ is the jet-resolution parameter which value
is set at convenience; in practice,
$0.4 < D < 1.0$. The algorithm is defined by means of an
iterative procedure. One starts with an empty list of jets and a list
of protojets, the latter being in the first step by definition identical
to the partons. Then, the quantities $d_i$ and $ d_{ij}$ are evaluated
for all the protojets and the minimum among them is found. If this
minimum is $d_i$, then protojet $i$ is moved from the list
of protojets to the list of jets. Otherwise, if the minimum is
$d_{ij}$, then protojets $i$ and $j$ are merged into a protojet.
The four-momentum of the protojet is defined by means of the merging
procedure of ref.~[\ref{D0report}]: it is the sum of the four-momenta
of the two constituent protojets. The jet-finding procedure
is repeated as long as there are protojets around. When the list of
jets is completed, that is, the list of protojets is empty, all jets
with $k_{\sss T}$ below a certain threshold are dropped. We choose the
threshold to be a given fraction (which we denote by $f$) of the
greatest squared transverse momentum; this means that the threshold
has to be recomputed for every event.

Other jet-finding algorithms are obviously
possible~[\ref{onejettheory},\ref{snowmass}]. Nevertheless,
preliminary studies~[\ref{D0report}] indicate that, contrary
to $e^+e^-$ annihilation, in hadroproduction the prescription
described here is favoured by the data.

\subsection{The measurement function}

Using the jet-finding algorithm of the previous section, we can
explicitly define the measurement functions we need to construct
the jet cross section.

When there are only three partons in the final state, no merging
is possible, and the partons themselves will eventually result
in physical jets. Therefore
\beqn
{\cal S}_3&=&\sum_{\sigma(J)}\delta(j,k,l)
\Th(min(d_j,d_k,d_l)-f\,max(d_j,d_k,d_l))
\nonumber \\*&&\,\,\times\,
\Th(R_{jk}-D^2)\Th(R_{jl}-D^2)\Th(R_{kl}-D^2)\,,
\label{S3def}
\eeqn
where we introduced the shorthand notation for the $\delta$
over four-momenta
\beq
\delta(j,k,l)=\delta(k_j-J_1)\delta(k_k-J_2)\delta(k_l-J_3).
\eeq
The indices $j$, $k$ and $l$ take the values $3,~4,~5$ and are different
from each other. $\sigma(J)$ denotes the permutation over the
jet four-momenta $\Jetlist$.

When four partons are present in the final state, the situation is
somewhat more involved. To get three jets starting from four
partons, only the following possibilities may occur:

$\bullet$ no merging, but one parton is dropped being below the hard
scale (this contribution will be denoted by ${\cal S}_i^{(0)}$);

$\bullet$ one merging, occurring in the first step of the algorithm
(${\cal S}_{ij}^{(1)}$);

$\bullet$ one merging, occurring in the second step of the algorithm
(${\cal S}_i^{(2)}$);

$\bullet$ one merging, occurring in the third step of the algorithm
(${\cal S}_i^{(3)}$).

\noindent
Consistently, we will then write
\beq
{\cal S}_4=\sum_i\left({\cal S}_i^{(0)}+{\cal S}_i^{(2)}
+{\cal S}_i^{(3)}\right)+\sum_{\stackrel{i,j}{i<j}}{\cal S}_{ij}^{(1)}\,,
\label{S40}
\eeq
where
\beqn
{\cal S}_i^{(0)}&=&\sum_{\sigma(J)}\delta(j,k,l){\cal F}_{jkl}^{(i,0)}\,,
\label{S4i1}
\\
{\cal S}_{ij}^{(1)}&=&\sum_{\sigma(J)}
\delta(i+j,k,l){\cal F}_{kl}^{(ij,1)}\,,
\label{S4ij3}
\\
{\cal S}_i^{(2)}&=&\sum_{\sigma(J_I)}\sum_{\stackrel{j,k,l}{k<l}}
\delta(i,j,k+l){\cal F}_{jkl}^{(i,2)}\,,
\\
{\cal S}_i^{(3)}&=&\sum_{\sigma(J_I)}\sum_{\stackrel{j,k,l}{k<l}}
\delta(i,j,k+l){\cal F}_{jkl}^{(i,3)}\,,
\eeqn
and the indices $j$, $k$, and $l$ can take the values
$3,\,4,\,5,\,6$ being different from each other and from $i$
\beq
\{j,k,l\}\,=\,\{3,\,4,\,5,\,6\}\,\backslash\,\{i\}\,.
\eeq
To write explicitly the ${\cal F}$ functions we introduce the
following shorthand definitions
\beqn
m(..,i,..,j+p,..,mn,..)=min(..,d_i,..,d_{j+p},..,d_{mn},..)\,,
\nonumber \\*
M(..,i,..,j+p,..,mn,..)=max(..,d_i,..,d_{j+p},..,d_{mn},..)\,,
\label{minmax}
\eeqn
and, consistently, we denote everywhere $d_i\equiv i$, $d_{ij}\equiv ij$.
We also write for short
\beq
m([a])=m(i,..,jk,..)\;\;\;\;\forall i\neq a,\;\;\forall jk\neq a,
\label{minbuta}
\eeq
where $d_a$ can be either $d_\alpha$ or $d_{\alpha\beta}$, and the
minimum on the RHS is evaluated over the list of all the $d_i$ and
$d_{jk}$ with the exclusion of $d_a$. Finally, we denote
\beq
m_{\{a,..,b,..\}}([c]),
\eeq
which is analogous to eq.~(\ref{minbuta}); the $\{a,..,b,..\}$
inserted means that the list of $d_i$ and $d_{ij}$ over which
the minimum is evaluated is such that it does not contain the
indices $a$ and $b$. We then have
\beqn
{\cal F}_{jkl}^{(i,0)}&=&\Th(m([i])-i)\Th(fM(j,k,l)-i)
\Th(m(j,k,l)-fM(j,k,l))
\nonumber \\*&\times&
\Th(R_{jk}-D^2)\Th(R_{jl}-D^2)\Th(R_{kl}-D^2)\,,
\label{F1}
\\
{\cal F}_{kl}^{(ij,1)}&=&\Th(m([ij])-ij)\Th(m(i+j,k,l)-fM(i+j,k,l))
\nonumber \\*&\times&
\Th(R_{i+j,k}-D^2)\Th(R_{i+j,l}-D^2)\Th(R_{kl}-D^2)\,,
\label{F3}
\\
{\cal F}_{jkl}^{(i,2)}&=&\Th(m([i])-i)\Th(m(i,j,k+l)-fM(i,j,k+l))
\nonumber \\*&\times&
\Th(m_{\{i\}}([kl])-kl)\Th(R_{k+l,j}-D^2)\,,
\label{F2b}
\\
{\cal F}_{jkl}^{(i,3)}&=&\Th(m([i])-i)\Th(m(i,j,k+l)-fM(i,j,k+l))
\nonumber \\*&\times&
\Th(m_{\{i\}}([j])-j)\Th(D^2-R_{kl})\,.
\label{F2a}
\eeqn

\subsection{Infrared singular regions}

It is well known that the perturbatively calculated QCD cross
sections, even after the ultraviolet renormalization, have a divergent
behaviour, arising from the regions in which a parton (either virtual,
that is, exchanged in a loop, or real, that is, emitted and contributing
to the final state kinematics) is soft or collinear to another parton.
In this section, we will investigate the behaviour of the
measurement function in this regions (which we will denote as
infrared singular regions) in the case when there are four partons
in the final state. The singular regions eventually occurring
are as follows

$\bullet$ parton $i$ ($i=3,\,4,\,5,\,6$) is soft ($k_i^0\,\to\,0$);

$\bullet$ parton $i$ is collinear to the incoming parton $1$ or $2$
($i\parallel 1$ or $i\parallel 2$);

$\bullet$ parton $i$ is collinear to parton $j$ ($i\parallel j$).

\noindent
{}From eqs.~(\ref{F1})-(\ref{F2a}), it is quite easy to prove
that the ${\cal F}$ functions are non vanishing only in
the following singular regions
\beqn
{\cal F}_{jkl}^{(i,0)}\;\;\;\;&\Rightarrow&\;\;\;\;
k_i^0\,\to\,0,\;\;i\parallel 1,\;\;i\parallel 2\,,
\label{list1}
\\
{\cal F}_{kl}^{(ij,1)}\;\;\;\;&\Rightarrow&\;\;\;\;
k_i^0\,\to\,0,\;\;k_j^0\,\to\,0,\;\;i\parallel j\,,
\label{list4}
\\
{\cal F}_{jkl}^{(i,2)}\;\;\;\;&\Rightarrow&\;\;\;\;
none\,,
\label{list2}
\\
{\cal F}_{jkl}^{(i,3)}\;\;\;\;&\Rightarrow&\;\;\;\;
none\,.
\label{list3}
\eeqn
Eq.~(\ref{list4}) suggests the following decomposition
\beq
{\cal S}_{ij}^{(1)}={\cal S}_{ij}^{(1)}\Th(d_j-d_i)
+{\cal S}_{ij}^{(1)}\Th(d_i-d_j)\,;
\label{Sij3dec}
\eeq
in this way, the first term in the RHS of eq.~(\ref{Sij3dec}) does not
get contributions when $j$ is soft, and the second one when $i$ is soft.
This in turn implies that, after some algebra, we can cast
eq.~(\ref{S40}) in the following form
\beq
{\cal S}_4=\sum_i \left({\cal S}_i^{(sing)} + {\cal S}_i^{(fin)}\right),
\label{S4fin}
\eeq
where
\beqn
{\cal S}_i^{(sing)}&=&{\cal S}_i^{(0)}+
\sum_j^{[i]}{\cal S}_{ij}^{(1)}\Th(d_j-d_i)\,,
\label{S4sing}
\\*
{\cal S}_i^{(fin)}&=&{\cal S}_i^{(2)}+{\cal S}_i^{(3)}\,,
\label{S4ns}
\eeqn
and the $[i]$ in the sum means that $j$ can take the values
$3,\,4,\,5,\,6$ with the exclusion of $i$.
The quantity ${\cal S}_i^{(fin)}$ does not get any contribution
from the singular regions; on the other hand, ${\cal S}_i^{(sing)}$
is different from 0 when $i$ is soft (or $i\parallel 1$, $i\parallel 2$),
but it is equal to zero when any other parton is soft (or
collinear to the incoming partons), thanks to the factor $\Th(d_j-d_i)$.
The region in which $i\parallel j$ contributes to ${\cal S}_i^{(sing)}$
and to ${\cal S}_j^{(sing)}$, but the $\Th(d_j-d_i)$ inserted in
eq.~(\ref{S4sing}) prevents any double counting.

We can finally investigate the form of the limiting behaviour
of the measurement function ${\cal S}_4$ in the infrared singular regions.

$\bullet$~$i$ is soft. From eqs.~(\ref{list1})-(\ref{list3})
and eq.~(\ref{S4fin}), we have
\beq
\lim_{k_i^0\to 0}{\cal S}_4=\lim_{k_i^0\to 0}\Bigg[
{\cal S}_i^{(0)}+\sum_j^{[i]}{\cal S}_{ij}^{(1)}\Bigg].
\label{splitlim}
\eeq
The limit of the ${\cal F}$ functions can be very easily evaluated
from their definition; it simply amounts to make the formal substitution
$i+j\to j$ wherever $i+j$ appears. We still need to do some
combinatorial algebra to get the limit of the full ${\cal S}_4$;
after writing explicitly the terms contributing to the
sum and using
\beq
\lim_{k_i^0\to 0}\Bigg[\Th(m([i])-i)+
\sum_j^{[i]}\Th(m([i j])-i j)\Bigg]=1
\eeq
we get
\beq
\lim_{k_i^0\to 0}{\cal S}_4={\cal S}_3([i])\,,
\label{S4sftlim}
\eeq
where
\beqn
{\cal S}_3([i])&=&\sum_{\sigma(J)}\delta(j,k,l)\Th(m(j,k,l)-f\,M(j,k,l))
\nonumber \\*&&\,\,\times\,
\Th(R_{jk}-D^2)\Th(R_{jl}-D^2)\Th(R_{kl}-D^2).
\eeqn
Notice that this quantity is identical to the ${\cal S}_3$ function
of eq.~(\ref{S3def}), but for the fact that the indices $j$, $k$
and $l$ can also take the value $6$ (and the value $i$ is excluded).
Eq.~(\ref{S4sftlim}) has an obvious physical meaning: when a parton
gets soft, the remaining partons act as physical jets. The measurement
function has then to coincide with the one defined for three partons
in the final state. We also point out that each term in the sum in
the RHS of eq.~(\ref{splitlim}) has a well defined soft limit:
\beqn
\lim_{k_i^0\to 0}{\cal S}_i^{(0)}&=&{\cal S}_3([i])
\Th(R_{ij}-D^2)\Th(R_{ik}-D^2)\Th(R_{il}-D^2),
\\
\lim_{k_i^0\to 0}{\cal S}_{ij}^{(1)}&=&{\cal S}_3([i])
\Th(D^2-R_{ij})\Th(R_{ik}-R_{ij})\Th(R_{il}-R_{ij}).
\eeqn

$\bullet$~$i\parallel 1$. From eqs.~(\ref{list1})-(\ref{list3})
and eq.~(\ref{S4fin}), we have
\beq
\lim_{\vec{k}_i\parallel\vec{k}_1}{\cal S}_4=
\lim_{\vec{k}_i\parallel\vec{k}_1}{\cal S}_i^{(0)}\,.
\eeq
It is quite easy to prove that, in this limit,
the following equation holds
\beq
\lim_{\vec{k}_i\parallel\vec{k}_1}\Th(m([i])-i)=1
\eeq
and therefore
\beq
\lim_{\vec{k}_i\parallel\vec{k}_1}{\cal S}_4={\cal S}_3([i])\,.
\label{S4clllim}
\eeq
The case in which $i\parallel 2$ is completely analogous and
gives an identical result.

$\bullet$~$i\parallel j$. From eqs.~(\ref{list1})-(\ref{list3})
and eq.~(\ref{S4fin}), we have
\beq
\lim_{\vec{k}_i\parallel\vec{k}_j}{\cal S}_4=
\lim_{\vec{k}_i\parallel\vec{k}_j}{\cal S}_{ij}^{(1)}\,,
\label{S4ijlimtmp}
\eeq
where we have used the fact that ${\cal S}_{ij}^{(1)}$ is symmetric
in the exchange of $i$ and $j$ and
\beq
\Th(d_i-d_j)+\Th(d_j-d_i)\equiv 1.
\eeq
In the limit at hand, $d_{ij}\to 0$ and therefore
\beq
\lim_{\vec{k}_i\parallel\vec{k}_j}\Th(m([ij])-ij)=1\,,
\eeq
while all the other $\Th$ functions containing $d_{ij}$
in the list over which the minimum is evaluated are zero.
Therefore we have
\beq
\lim_{\vec{k}_i\parallel\vec{k}_j}{\cal S}_4=
{\cal S}_3([ij])\,,
\label{S4albelim}
\eeq
where
\beqn
{\cal S}_3([ij])&=&\sum_{\sigma(J)}\delta(p,k,l)\Th(m(p,k,l)-f\,M(p,k,l))
\nonumber \\*&&\,\,\times\,
\Th(R_{pk}-D^2)\Th(R_{pl}-D^2)\Th(R_{kl}-D^2).
\eeqn
and the indices $p$, $k$ and $l$ in the sum can now take also
the value $7$, having defined $k_7=k_i+k_j$
(notice that this formal manipulation allows to maintain the
notation used in the previous case; the meaning of eq.(\ref{S4albelim})
is that the final state collinear limit of the ${\cal S}_4$
function is again a jet-defining function of three partons into
three jets, where one of the partons has four-momentum equal to
the sum of the four-momenta of the partons becoming collinear).

We finally point out that in the numerical implementation of
the algorithm we will also need to consider the case where one
parton is soft {\it and} collinear to an incoming parton or to a final
state parton. In this case, we find
\beqn
\lim_{\vec{k}_i\parallel\vec{k}_1}
\lim_{k_i^0\to 0}{\cal S}_i^{(0)}&=&{\cal S}_3([i]),
\label{softcolllim1}
\\
\lim_{\vec{k}_i\parallel\vec{k}_j}
\lim_{k_i^0\to 0}{\cal S}_{ij}^{(1)}&=&{\cal S}_3([ij]).
\label{softcolllim2}
\eeqn
Notice that in eqs.~(\ref{softcolllim1}) and~(\ref{softcolllim2})
the order in which the limits are taken is irrelevant. Also,
eq.~(\ref{softcolllim1}) holds true when $i\parallel 2$.

In the following, we will use the properties of the measurement
function ${\cal S}_4$ to disentangle the structure of the
singularities in the real contribution. In spite of this fact,
our method is completely general. In fact, we will basically
rely only upon eqs.~(\ref{S4fin})-(\ref{S4ns}), that is, on the
decomposition of the measurement function into terms which get
contribution from two singular regions at the worst. It should
be clear that such a decomposition can always be performed, being
essentially due to the infrared safeness requirement on the
measurement function.

\subsection{Organization of the calculation}

Given the definitions of the measurement functions and of the collinear
counterterms, the main problem we must tackle now is to arrange for
all the singularities appearing in the perturbatively calculated
cross sections to cancel analytically. Notice that
this cancellation is achieved only after that the real and the virtual
contributions are summed together, as in eq.~(\ref{realplusvirt}).
Also, it must take place already at the partonic level, that is, we do
not need to take into account the convolution with the partonic densities
in the intermediate steps of the calculation (to have a physical
picture of this fact, one can imagine to have very peaked partonic
distribution functions, becoming a $\delta(1-x)$ in some limit).
A possible choice for the reference frame in which the calculation
is carried out is that of the partonic center-of-mass one; this
results in a simplification of the kinematics.
We begin by considering the singular structure of the cross section;
it is apparent from their definition that the measurement
functions are boost invariant; therefore, proving the cancellation of
the singularities in the partonic center-of-mass frame, ensures that
the same cancellation holds true also in the hadronic center-of-mass frame.
After the cancellation of singularities, we are left with a subtracted
partonic cross section, analogous to $d\hat{\sigma}_{ab}$ of
eq.~(\ref{factth}), defined in the partonic center-of-mass frame
instead of the hadronic center-of-mass frame. Clearly, the two are
related by a longitudinal boost. The simplest way to perform
this boost is by numerical methods (we stress that there is no loss of
generality: in all practical cases, the integration in eq.~(\ref{factth})
has to be carried out numerically).

Since the real and virtual part of the cross section are both
divergent, we have to write them in a form
in which the structure of the divergencies is very clearly displayed.
To this end, the most involved case is that of the real contribution.
Taking into account eq.~(\ref{S4fin}), we split the real part of the
cross section as follows
\beq
d\sigma_{\aoat}^{(r)}=\sum_{i}\left(d\sigma_{\aoat,i}^{(sing)}
+d\sigma_{\aoat,i}^{(fin)}\right),
\label{sigrealsplit}
\eeq
where
\beqn
d\sigma_{\aoat,i}^{(fin)}&=&
\frac{1}{4!}\FLFsum\MF(\FLfull) {\cal S}_i^{(fin)} d\phi_4\,,
\label{sigrns}
\\
d\sigma_{\aoat,i}^{(sing)}&=&
\frac{1}{4!}\FLFsum\MF(\FLfull) {\cal S}_i^{(sing)} d\phi_4\,.
\label{sigrsing}
\eeqn
$d\sigma_{\aoat,i}^{(fin)}$ in eq.~(\ref{sigrns}) is finite, and it is
already suited for numerical computations. On the other hand, all the
singular contributions are contained in $d\sigma_{\aoat,i}^{(sing)}$,
eq.~(\ref{sigrsing}). The form of ${\cal S}_i^{(sing)}$ implies
that $d\sigma_{\aoat,i}^{(sing)}$ has the following singularities:
$i$ soft, $i\parallel 1$, $i\parallel 2$ and $i\parallel j$,
$\forall j\neq i$. From the previous discussion, we know that the
singularities due to initial state collinear emission are cancelled
by the addition of the collinear counterterms, eqs.~(\ref{cnt1})
and~(\ref{cnt2}). It is then useful to define a subtracted
real contribution in the following way
\beq
d\hat{\sigma}_{\aoat}^{(r)}=d\sigma_{\aoat}^{(r)}
+d\sigma_{\aoat}^{(cnt+)}+d\sigma_{\aoat}^{(cnt-)}
\eeq
that is
\beq
d\hat{\sigma}_{\aoat}^{(r)}=\sum_{i}\left(
d\hat{\sigma}_{\aoat,i}^{(sing)}+d\sigma_{\aoat,i}^{(fin)}\right)\,,
\eeq
where
\beq
d\hat{\sigma}_{\aoat,i}^{(sing)}=d\sigma_{\aoat,i}^{(sing)}
+\frac{1}{4}d\sigma_{\aoat}^{(cnt+)}
+\frac{1}{4}d\sigma_{\aoat}^{(cnt-)}\,.
\label{dsighatreal}
\eeq
The key point here is that no parton but $i$
can be soft. This amounts to a very clean disentangling of the
soft regions, and allows to factor out immediately the pure soft
singularities, as we will show later.

In the following sections, we will first summarize the results
for the virtual contribution, known from the literature. Next, we
will discuss the separation of the soft and collinear singularities
in the real contribution. We will then be able to analytically evaluate
the structure of these singularities. Finally, we will collect
the results in a form suitable for numerical evaluation.
\section{Virtual contribution}

Although the calculation of the loop corrections to the two-to-three
partonic subprocesses is quite involved, and the result is
rather complicated, the structure of the divergent terms
is indeed very simple (see e.g.
refs.~[\ref{kssub},\ref{KSTsing},\ref{ESscale}])
\beq
\MTo(\ArgTfull)=\frac{\as}{2\pi}\SVfact\,\FLTsum\,{\cal V}(\ArgTfull)\,,
\label{virttransamp}
\eeq
where
\beqn
{\cal V}(\ArgTfull)&=&
-\Bigg(\frac{1}{\ep^2}\sum_{n=1}^{5}C(a_n)
+\frac{1}{\ep}\sum_{n=1}^{5}\gamma(a_n)\Bigg)\MTz(\ArgTfull)
\nonumber \\*&&
+\frac{1}{2\ep}\sum_{\stackrel{n,m=1}{n\neq m}}^{5}
\log\frac{2k_n\cdot k_m}{Q^2}\frac{1}{8\pi^2}\MTmn(\ArgTfull)
\nonumber \\*&&
+\MTo_{\sss NS}(\ArgTfull)\,.
\label{virtstruct}
\eeqn
In these equations, $\mu$ is the renormalization scale
and $Q$ is an arbitrary mass scale, introduced by Ellis
and Sexton in ref.~[\ref{ESscale}] to facilitate the
writing of the result.
The quantities $\MTo$ and $\MTz$ were defined respectively
in eqs.~(\ref{virtampdef}) and~(\ref{bornampdef}); the
$\MTmn$ are usually denoted as color-linked Born squared
amplitudes. They are symmetric in $m,n$ and satisfy the identity
\beq
\sum_{\stackrel{n=1}{n\neq m}}^{5}\MTmn(\ArgTfull)=
16\,C(a_m)\,\pi^2\MTz(\ArgTfull)\,;
\label{mmnident}
\eeq
the reason for the $1/8\pi^2$ normalization factor (that can be
freely chosen, provided that eq.~(\ref{mmnident}) is suitably modified)
inserted in front of these terms in eq.~(\ref{virtstruct}) will become
clear in the following. The factor
\beq
\frac{(4\pi)^\ep}{\Gamma(1-\ep)}=
\frac{\Gamma(1+\ep)\Gamma(1-\ep)^2}{\Gamma(1-2\ep)}+{\cal O}(\ep^3)
\eeq
naturally results from loop integration. In eq.~(\ref{virtstruct})
we also made use of the flavour dependent quantities $C(a_n)$
and $\gamma(a_n)$; for $SU(N_c)$ color group, they are
\beqn
C(g)=\CA=N_c\,,\phantom{aaaa}&&\;\;\;\;
\gamma(g)=\frac{11\CA-4\TF N_f}{6}\,,
\label{gCasimir}
\\*
C(q)=\CF=\frac{N_c^2-1}{2N_c}\,,&&\;\;\;\;\gamma(q)=\frac{3}{2}\CF\,,
\label{qCasimir}
\eeqn
where $\TF=1/2$ and $N_f$ is the number of quark flavours.
All the non-divergent terms in eq.~(\ref{virtstruct}) were
collected in $\MTo_{\sss NS}$. The explicit results can be found
in ref.~[\ref{BDK5g}] for 5g process, ref.~[\ref{KST4q1g}] for 4q1g
process, ref.~[\ref{BDK2q3g},\ref{ASthesis}] for 3g2q process;
the calculations were carried out using helicity amplitude methods
in the dimensional reduction (DR) scheme.
In the DR scheme, $\MTz$ and $\MTmn$ are evaluated in four dimensions,
while in the conventional dimensional regularization scheme (CDR)
they are in $4-2\ep$ dimensions. As pointed out in ref.~[\ref{KST2to2}],
to convert the results obtained in the DR
scheme~[\ref{BDK5g}-\ref{ASthesis}] into the CDR scheme, one has
to use $\MTz$ and $\MTmn$ in $4-2\ep$ dimensions instead of four
dimensions; furthermore, the finite part is modified as follows
\beq
\MTo_{\sss NS}({\scriptstyle\rm CDR})=\MTo_{\sss NS}({\scriptstyle\rm DR})
-\MTz\,\sum_{n=1}^5\,\tilde{\gamma}(a_n)\,,
\eeq
where the universal factors $\tilde{\gamma}$ are
\beq
\tilde{\gamma}(g)=\frac{N_c}{6}\,,\;\;\;\;
\tilde{\gamma}(q)=\frac{1}{2}\frac{N_c^2-1}{2N_c}\,.
\eeq
\section{Real contribution}

\subsection{Separation of singularities}

We now turn to the real contribution, defined in eq.~(\ref{realjetdef})
and further decomposed in eq.~(\ref{sigrealsplit}). Thanks
to the fact that no divergencies are present in eq.~(\ref{sigrns}),
we will deal only with eq.~(\ref{sigrsing}). In the partonic
center-of-mass frame, the incoming partons have momentum
\beqn
k_1&=&\frac{\sqrt{S}}{2}(1,\vec{0},1),
\\*
k_2&=&\frac{\sqrt{S}}{2}(1,\vec{0},-1),
\eeqn
where $\sqrt{S}$ is the partonic center-of-mass energy and $\vec{0}$ is
the null vector in a $(2-2\ep)$-dimensional space. In this frame, we write the
momentum of parton $i$ as
\beq
k_i=\frac{\sqrt{S}}{2}\xi_i\left(1,\sqrt{1-y_i^2}
\vec{e}_{i{\sss T}},y_i\right),
\label{kidef}
\eeq
where $\vec{e}_{i{\sss T}}$ is a unit vector in the $(2-2\ep)$-dimensional
transverse momentum space, $-1\leq y_i\leq 1$ and $0\leq\xi_i\leq 1$.
By construction, when $\xi_i\to 0$ the parton $i$ gets soft, and when
$y_i\to\pm 1$ it gets collinear to the incoming partons.
With this parametrization, in $4-2\ep$ dimensions the invariant measure
over the variables of parton $i$ is
\beq
d\phi(i)=\frac{d^{3-2\ep}k_i}{(2\pi)^{3-2\ep}\,2k_i^0}=
\frac{1}{2(2\pi)^{3-2\ep}}
\left(\frac{\sqrt{S}}{2}\right)^{2-2\ep}\xi_i^{1-2\ep}
\left(1-y_i^2\right)^{-\ep} d\xi_i dy_i d\Omega_i^{(2-2\ep)}\,,
\label{dphii}
\eeq
where $d\Omega_i^{(2-2\ep)}$ is the angular measure in $2-2\ep$ dimensions.
We also write the four-body phase space as $d\phi_4=d\phi d\phi(i)$, where
\beq
d\phi=(2\pi)^{4-2\ep}\delta^{4-2\ep}\left(k_1+k_2-\sum_{l=3}^6 k_l\right)
\prod_l^{[i]}\frac{d^{3-2\ep}k_l}{(2\pi)^{3-2\ep}\,2k_l^0}\,.
\label{dphidef}
\eeq
We now regulate the soft singularities by multiplying the
invariant amplitude squared by $\xi_i^2$.
Taking into account eq.~(\ref{dphii}), eq.~(\ref{sigrsing}) becomes
\beqn
d\sigma_{\aoat,i}^{(sing)}&=&\frac{\xi_i^2}{4!}\FLFsum
\MF(\FLfull)\,{\cal S}_i^{(sing)} d\phi
\nonumber \\*&\times&
\frac{1}{2(2\pi)^{3-2\ep}}\left(\frac{\sqrt{S}}{2}\right)^{2-2\ep}
\xi_i^{-1-2\ep} \left(1-y_i^2\right)^{-\ep}
d\xi_i dy_i d\Omega_i^{(2-2\ep)}\,.
\label{3jetreg}
\eeqn
We can make use of the following identity
\beq
\xi_i^{-1-2\ep}=-\frac{\xi_{cut}^{-2\ep}}{2\ep}\delta(\xi_i)
+\left(\frac{1}{\xi_i}\right)_c -2\ep\left(\frac{\log\xi_i}{\xi_i}\right)_c
+ {\cal O}(\ep^2)\,,
\label{xiid}
\eeq
where $\xi_{cut}$ is an arbitrary parameter satisfying the
condition $0<\xi_{cut}\leq 1$, and the distributions in the
RHS are defined as follows
\beqn
<\uoxiic,f>&=&\int_0^1 d\xi_i
\frac{f(\xi_i)-f(0)\Th(\xi_{cut}-\xi_i)}{\xi_i}\,,
\label{uoxidef}
\\
<\uoxiilc,f>&=&\int_0^1 d\xi_i
\Bigg[f(\xi_i)-f(0)\Th(\xi_{cut}-\xi_i)\Bigg]\frac{\log\xi_i}{\xi_i}\,.
\eeqn
Notice that different $\xi_{cut}$ can be chosen for different $i$; here
we restricted to the simplest case. Substituting eq.~(\ref{xiid})
into eq.~(\ref{3jetreg}) we get
\beq
d\sigma_{\aoat,i}^{(sing)}=d\sigma_{\aoat,i}^{(s)}
+d\sigma_{\aoat,i}^{(ns)}\,,
\label{sigsingsplit}
\eeq
where
\beqn
d\sigma_{\aoat,i}^{(s)}&=&-\frac{\xi_{cut}^{-2\ep}}{2\ep}\delta(\xi_i)d\xi_i
\left(\frac{\xi_i^2}{4!}\FLFsum\MF(\FLfull)\right)\,
{\cal S}_i^{(sing)} d\phi
\nonumber \\*&\times&
\frac{1}{2(2\pi)^{3-2\ep}}\left(\frac{\sqrt{S}}{2}\right)^{2-2\ep}
\left(1-y_i^2\right)^{-\ep} dy_i d\Omega_i^{(2-2\ep)}\,,
\label{sigsoft}
\\
d\sigma_{\aoat,i}^{(ns)}&=&\Bigg[\uoxiic-2\ep\uoxiilc\Bigg]
\left(\frac{\xi_i^2}{4!}\FLFsum\MF(\FLfull)\right)\,
{\cal S}_i^{(sing)} d\phi
\nonumber \\*&\times&
\frac{1}{2(2\pi)^{3-2\ep}}\left(\frac{\sqrt{S}}{2}\right)^{2-2\ep}
\left(1-y_i^2\right)^{-\ep} d\xi_i dy_i d\Omega_i^{(2-2\ep)}\,.
\label{signsoft}
\eeqn
Eq.~(\ref{sigsoft}) does contain the singularities due to parton
$i$ becoming soft, while eq.~(\ref{signsoft}) is free of them,
but does contain collinear singularities. We can also fully disentangle
the collinear singularities of $d\sigma_{\aoat,i}^{(ns)}$.
Using eq.~(\ref{list1}) and eq.~(\ref{list4}) we notice that
a single term in the ${\cal S}_i^{(sing)}$ function gets contributions
from one singular collinear region at the worst.
Therefore, using eq.~(\ref{S4sing}), we can naturally split
$d\sigma_{\aoat,i}^{(ns)}$ into several terms:
\beq
d\sigma_{\aoat,i}^{(ns)}=d\sigma_{\aoat,i}^{(in)}
+\sum_{j}^{[i]}d\sigma_{\aoat,ij}^{(out)}\,,
\label{signssplit}
\eeq
where
\beqn
d\sigma_{\aoat,i}^{(in)}&=&\Bigg[\uoxiic-2\ep\uoxiilc\Bigg]
\left(\frac{\xi_i^2}{4!}\FLFsum\MF(\FLfull)\right)
{\cal S}_i^{(0)} d\phi
\nonumber \\*&\times&
\frac{1}{2(2\pi)^{3-2\ep}}\left(\frac{\sqrt{S}}{2}\right)^{2-2\ep}
\left(1-y_i^2\right)^{-\ep} d\xi_i dy_i d\Omega_i^{(2-2\ep)}\,,
\label{sigiin}
\\
d\sigma_{\aoat,ij}^{(out)}&=&\Bigg[\uoxiic-2\ep\uoxiilc\Bigg]
\left(\frac{\xi_i^2}{4!}\FLFsum\MF(\FLfull)\right)
{\cal S}_{ij}^{(1)} \Th(d_j-d_i) d\phi
\nonumber \\*&\times&
\frac{1}{2(2\pi)^{3-2\ep}}\left(\frac{\sqrt{S}}{2}\right)^{2-2\ep}
\left(1-y_i^2\right)^{-\ep} d\xi_i dy_i d\Omega_i^{(2-2\ep)}\,.
\label{sigijout}
\eeqn
By construction, eq.~(\ref{sigiin}) contains only initial state
collinear singularities, while eq.~(\ref{sigijout}) contains only
final state ones.

In summary, the decompositions defined in eqs.~(\ref{sigrealsplit}),
(\ref{sigsingsplit}) and~(\ref{signssplit}) allow to write the real
contribution to the partonic cross section in the following way
\beq
d\sigma_{\aoat}^{(r)}=\sum_{i}\left(d\sigma_{\aoat,i}^{(fin)}
+d\sigma_{\aoat,i}^{(s)}+d\sigma_{\aoat,i}^{(in)}
+\sum_{j}^{[i]}d\sigma_{\aoat,ij}^{(out)}\right),
\label{sigrealsplit2}
\eeq
where the quantities in the RHS of this equation have been presented
in eqs.~(\ref{sigrns}), (\ref{sigsoft}), (\ref{sigiin}),
and~(\ref{sigijout}) respectively. As far as the real contribution
is concerned, eq.~(\ref{sigrealsplit2}) will be regarded as our master
equation. The key point is that every term in the RHS gets contributions
from two singular regions at the worst. In the next three subsections,
we will calculate the singular contributions explicitly.
At the very end, we will show that, as expected, the sum of the real
contribution, of the virtual contribution and of the collinear
counterterms is finite.
\subsection{Soft singularities}

We begin by considering eq.~(\ref{sigsoft}). The $\delta(\xi_i)$
allows to take the soft limit of all the quantities appearing in
that equation. In particular, we have
\beq
\lim_{\xi_i\to 0}\MF(\FLfull;\SFfull)=
\delta_{ga_i}\frac{\as \mu^{2\ep}}{2\pi}
\sum_{\stackrel{n,m}{n<m}}^{[i]}
\frac{k_n\cdot k_m}{k_n\cdot k_i~k_m\cdot k_i}
\MTmn\left(\FLfullbi;\SFfullbi\right),
\label{ttfsoft}
\eeq
where $n$ and $m$ in the sum run also over the values $1$ and $2$
(incoming partons).
In eq.~(\ref{ttfsoft}) we inserted $\mu^{2\ep}$ as customary
in $4-2\ep$ dimensions. Eq.~(\ref{ttfsoft}) also explains the reason
for the normalization of the $\MTmn$ terms in eq.~(\ref{virtstruct});
we choose to keep the form of the soft limit as simple as possible.
Taking into account eq.~(\ref{S4sftlim}) and the formal limit
\beq
\lim_{\xi_i\to 0}d\phi=d\phi_3\left(k_1,k_2\to\SFjbi\right),
\eeq
and using $\FLFsum\delta_{ga_i}=\FLFsumbi$,
eq.~(\ref{sigsoft}) becomes therefore
\beqn
d\sigma_{\aoat,i}^{(s)}&=&-\frac{\as}{2\pi}\frac{\xi_{cut}^{-2\ep}}{2\ep}
\frac{2^{2\ep}}{2(2\pi)^{3-2\ep}}\left(\frac{S}{\mu^2}\right)^{-\ep}
\nonumber \\*&\times&
\sum_{\stackrel{n,m}{n<m}}^{[i]}
\uoffct\FLFsumbi\MTmn\left(\FLfullbi;\SFfullbi\right)\,
{\cal S}_3([i])\, d\phi_3\left(k_1,k_2\to\SFjbi\right)
\nonumber \\*&\times&
\delta(\xi_i)\left(\frac{\sqrt{S}}{2}\right)^2
\frac{k_n\cdot k_m}{k_n\cdot k_i~k_m\cdot k_i}
\xi_i^2\left(1-y_i^2\right)^{-\ep} d\xi_i dy_i d\Omega_i^{(2-2\ep)}\,.
\label{sigsoft2}
\eeqn
The dependence of the RHS of eq.~(\ref{sigsoft2}) upon the
variables $y_i$ and $\Omega_i$ is fully contained in the last
line. The integral over $d\xi_i$, $d y_i$ and $d\Omega_i^{(2-2\ep)}$
can therefore be performed explicitly; the result is reported
in appendix A. We can cast eq.~(\ref{sigsoft2}) in the following form
\beqn
d\sigma_{\aoat,i}^{(s)}&=&\frac{\as}{2\pi}
\sum_{\stackrel{n,m}{n<m}}^{[i]}
\left({\cal I}_{mn}^{(div)}+{\cal I}_{mn}^{(reg)}\right)
\nonumber \\*&\times&
\uoffct\FLFsumbi\MTmn\left(\FLfullbi;\SFfullbi\right)\,
{\cal S}_3([i])\, d\phi_3\left(k_1,k_2\to\SFjbi\right),
\phantom{aaa}
\label{softivsmn}
\eeqn
with
\beq
{\cal I}_{mn}^{(div)}+{\cal I}_{mn}^{(reg)}=
-\frac{\xi_{cut}^{-2\ep}}{2\ep}\frac{2^{2\ep}}{2(2\pi)^{3-2\ep}}
\left(\frac{S}{\mu^2}\right)^{-\ep} {\cal J}_{mn}\,,
\eeq
where ${\cal J}_{mn}$ is given in eq.~(\ref{Icnlnm}). Explicitly
\beqn
{\cal I}_{mn}^{(div)}&=&\frac{1}{8\pi^2}\SVfact\Bigg[\frac{1}{\ep^2}
-\frac{1}{\ep}\left(\log\frac{2k_n\cdot k_m}{Q^2}
-\log\frac{4E_n E_m}{\xi_{cut}^2 S}\right)\Bigg],
\label{Imndiv}
\\
{\cal I}_{mn}^{(reg)}&=&\frac{1}{8\pi^2}\Bigg[
\frac{1}{2}\log^2\frac{\xi_{cut}^2 S}{Q^2}
+\log\frac{\xi_{cut}^2 S}{Q^2}\log\frac{k_n\cdot k_m}{2E_n E_m}
-{\rm Li}_2\left(\frac{k_n\cdot k_m}{2E_n E_m}\right)
\nonumber \\*&&\phantom{\frac{1}{8\pi^2}}
+\frac{1}{2}\log^2\frac{2k_n\cdot k_m}{E_n E_m}
-\log\left(4-\frac{2 k_n\cdot k_m}{E_n E_m}\right)
\log\frac{k_n\cdot k_m}{2E_n E_m}-2\log^2 2\Bigg].\phantom{aaaaa}
\label{Imnreg}
\eeqn
Here $E_n$ is the energy of the parton $n$ in the partonic
center-of-mass frame and $Q$ is the mass scale introduced in
eq.~(\ref{virtstruct}) . We can express $E_n$ in term of invariant
quantities through the equation
\beq
E_n=\frac{k_1\cdot k_n + k_2\cdot k_n}{\sqrt{S}}\,.
\eeq
It is apparent that the dependence upon the index $i$ in
eq.~(\ref{softivsmn}) is immaterial. We can therefore relabel
the partons at will and get
\beq
d\sigma_{\aoat}^{(s)}=\sum_i d\sigma_{\aoat,i}^{(s)}
=\frac{\as}{2\pi}\,\frac{1}{2}
\sum_{\stackrel{n,m=1}{n\neq m}}^{5}
\left({\cal I}_{mn}^{(div)}+{\cal I}_{mn}^{(reg)}\right)
\FLTsum d\sigma_{mn}^{(0)}(\FLTfull)
\label{softivsmn2}
\eeq
where we have set
\beq
d\sigma_{mn}^{(0)}\left(\ArgTfull\right)=
\uotfct\MTmn\left(\ArgTfull\right)\,
{\cal S}_3\,d\phi_3\left(k_1,k_2\to\STj\right)
\label{bornmndef}
\eeq
(in the following, as in eq.~(\ref{softivsmn2}), we will not
indicate explicitly the dependence upon the momenta of this quantity).
Exploiting eq.~(\ref{mmnident}) and using the fact
that $E_1=E_2=\sqrt{S}/2$, eq.~(\ref{softivsmn2}) becomes
\beqn
d\sigma_{\aoat}^{(s)}&=&\frac{\as}{2\pi}\SVfact\FLTsum
\Bigg[\frac{1}{\ep^2}\sum_{n=1}^{5}C(a_n)
+\frac{2}{\ep}\sum_{j=3}^{5}C(a_j)\log\frac{2E_j}{\xi_{cut}\sqrt{S}}
\nonumber \\*&&\phantom{a}
-\frac{2}{\ep}\left(C(a_1)+C(a_2)\right)\log\xi_{cut}\Bigg]
d\sigma^{(0)}(\FLTfull)
\nonumber \\*&-&
\frac{\as}{2\pi}\SVfact\,\frac{1}{2\ep}
\sum_{\stackrel{n,m=1}{n\neq m}}^{5}
\log\frac{2k_n\cdot k_m}{Q^2}\,\frac{1}{8\pi^2}\,
\FLTsum d\sigma_{mn}^{(0)}(\FLTfull)
\nonumber \\*&+&
\frac{\as}{2\pi}\,\frac{1}{2}\,
\sum_{\stackrel{n,m=1}{n\neq m}}^{5}
{\cal I}_{mn}^{(reg)}\FLTsum d\sigma_{mn}^{(0)}(\FLTfull)\,.
\label{softfact}
\eeqn
where, consistently with eq.~(\ref{bornmndef}),
\beq
d\sigma^{(0)}(\ArgTfull)=
\uotfct\MTz\left(\ArgTfull\right)\,
{\cal S}_3\,d\phi_3\left(k_1,k_2\to\STj\right);
\label{bornfldef}
\eeq
this quantity is just the Born cross section of eq.~(\ref{bornjetdef})
without the sum over the flavours of the final state partons.
We point out that the second and the third singular term in
eq.~(\ref{softfact}) are cancelled by similar terms appearing in the final
and initial state contributions. The remaining singular terms are
cancelled by the virtual contribution.
\subsection{Initial state singularities}

We now turn to eq.~(\ref{sigiin}).
The divergencies due to parton $i$ becoming collinear to one of the
incoming partons can simultaneously be regulated by multiplying the invariant
amplitude squared by the factor $(1-y_i^2)$. Eq.~(\ref{sigiin}) becomes
\beqn
d\sigma_{\aoat,i}^{(in)}&=&\uoffct\Bigg[\uoxiic
-2\ep\uoxiilc\Bigg]\left((1-y_i^2)\xi_i^2
\FLFsum\MF(\FLfull)\right) {\cal S}_i^{(0)} d\phi
\nonumber \\*&\times&
\frac{1}{2(2\pi)^{3-2\ep}}\left(\frac{\sqrt{S}}{2}\right)^{2-2\ep}
\left(1-y_i^2\right)^{-1-\ep} d\xi_i dy_i d\Omega_i^{(2-2\ep)}\,.
\label{sigiin2}
\eeqn
We use the identity
\beq
\left(1-y_i^2\right)^{-1-\ep}=-\frac{(2\delta_{\sss I})^{-\ep}}{2\ep}
\Bigg[\delta(1-y_i)+\delta(1+y_i)\Bigg]+{\cal P}(y_i)+{\cal O}(\ep),
\label{yid}
\eeq
where
\beq
{\cal P}(y_i)=\frac{1}{2}\Bigg[\uoyimdi+\uoyipdi\Bigg]\,.
\eeq
Here $\delta_{\sss I}$ is an arbitrary parameter satisfying the condition
$0<\delta_{\sss I}\leq 2$, and we defined
\beqn
<\uoyimdi,f>&=&\int_{-1}^{1}dy_i\,
\frac{f(y_i)-f(1)\Th(y_i-1+\delta_{\sss I})}{1-y_i}\,,
\\
<\uoyipdi,f>&=&\int_{-1}^{1}dy_i\,
\frac{f(y_i)-f(-1)\Th(-y_i-1+\delta_{\sss I})}{1+y_i}\,.
\eeqn
Very much like the $\xi_{cut}$ parameter, it would be possible to
choose a different $\delta_{\sss I}$ for every $i$; we prefer
however to avoid the proliferation of such parameters.
Eq.~(\ref{sigiin2}) can be splitted into three terms, the first two which
contain the collinear singularities (and in which the invariant amplitude
is substituted with its collinear limit), and the third one which is
finite in the limits $y_i\to\pm 1$. Explicitly
\beq
d\sigma_{\aoat,i}^{(in)}=d\sigma_{\aoat,i}^{(in,+)}
+d\sigma_{\aoat,i}^{(in,-)}+d\sigma_{\aoat,i}^{(in,f)}\,,
\label{sigiinsplit}
\eeq
where
\beqn
d\sigma_{\aoat,i}^{(in,f)}&=&\uoffct {\cal P}(y_i)
\Bigg[\uoxiic-2\ep\uoxiilc\Bigg]
\frac{1}{2(2\pi)^{3-2\ep}}\left(\frac{\sqrt{S}}{2}\right)^{2-2\ep}
\nonumber \\*&\times&
\left((1-y_i^2)\xi_i^2\FLFsum\MF(\FLfull)\right)
{\cal S}_i^{(0)} d\phi d\xi_i dy_i d\Omega_i^{(2-2\ep)}\,,
\label{sigiinfin}
\eeqn
and
\beqn
d\sigma_{\aoat,i}^{(in,\pm)}&=&-\frac{(2\delta_{\sss I})^{-\ep}}{2\ep}
\delta(1\mp y_i)\uoffct\Bigg[\uoxiic-2\ep\uoxiilc\Bigg]
\frac{1}{2(2\pi)^{3-2\ep}}\left(\frac{\sqrt{S}}{2}\right)^{2-2\ep}
\nonumber \\*&\times&
\left((1-y_i^2)\xi_i^2\FLFsum\MF(\FLfull)\right)
{\cal S}_i^{(0)} d\phi d\xi_i dy_i d\Omega_i^{(2-2\ep)}\,.
\label{sigcpm}
\eeqn
Thanks to the subtraction prescriptions of ${\cal P}(y_i)$,
which regulate initial state divergencies, and to eq.~(\ref{list1}),
which guarantees that ${\cal S}_i^{(0)}$ does not get contributions
from final state collinear divergencies, eq.~(\ref{sigiinfin}) is finite,
and therefore we can set $\ep=0$. We get
\beqn
d\sigma_{\aoat,i}^{(in,f)}&=&\uoffct\frac{1}{2}\uoxiic
\Bigg[\uoyimdi+\uoyipdi\Bigg]
\frac{1}{2(2\pi)^{3}}\left(\frac{\sqrt{S}}{2}\right)^{2}
\nonumber \\*&\times&
\left((1-y_i^2)\xi_i^2\FLFsum\MF(\FLfull)\right)
{\cal S}_i^{(0)} d\phi d\xi_i dy_i d\varphi_i\,.
\label{sigiinfin2}
\eeqn
Since all the divergencies have been properly regulated, the
RHS of eq.~(\ref{sigiinfin2}) can be numerically integrated.

We still have to deal with the divergent part of eq.~(\ref{sigiinsplit}),
that is to say, with the quantities defined in eq.~(\ref{sigcpm}).
The $\delta(1\mp y_i)$ in eq.~(\ref{sigcpm}) allows to take the
appropriate collinear limit of all the quantities appearing in
that equation.
We only consider $d\sigma_{\aoat,i}^{(in,+)}$, that is the
case in which $i\parallel 1$; the treatment of $i\parallel 2$ is
completely analogous. The collinear limit of the invariant amplitude
squared can be written as (see appendix B)
\beqn
&&\lim_{y_i\to 1}\MF(\FLfull;\SFfull)=
\frac{4\pi\as\mu^{2\ep}}{k_i\cdot k_1}
\Delta\left(\FLfull;\SFfull\right)
\phantom{\FLjbi;zk_1,k_2,\SFjbi aaa}
\nonumber \\*&&\phantom{\lim_{\vec{k}_i\parallel\vec{k}_1}}
+\frac{4\pi\as\mu^{2\ep}}{k_i\cdot k_1}
P_{S(a_1,\bar{a}_i)a_1}^{<}(z,\ep)\MTz
\left(S(a_1,\bar{a}_i),a_2,\FLjbi;zk_1,k_2,\SFjbi\right),
\label{limcoll}
\eeqn
where $P_{ab}^{<}(z,\ep)$ is the Altarelli-Parisi kernel for $z<1$
in $4-2\ep$ dimensions and $S(c,d)$ is the flavour of the parton
which can split into two partons of flavour $c$ and $d$;
for example, $S(g,g)=g$ and $S(q,g)=q$; by definition, if the
splitting into the flavours $c$ and $d$ is not possible,
$P_{S(c,d)a}^{<}$ is zero.
The $z$ parameter is such that, in the collinear limit,
$k_i=(1-z)k_1$. Taking into account eq.~(\ref{kidef}) we have
\beq
\xi_i\equiv 1-z\;\;\;\;{\rm if}\;\;\;\; y_i=1\,.
\label{xieqz}
\eeq
Using eq.~(\ref{limcoll}), we can then write
\beqn
&&\lim_{y_i\to 1}\left(1-y_i^2\right)\xi_i^2\MF(\FLfull;\SFfull)=
8\pi\as\mu^{2\ep}\left(\frac{2}{\sqrt{S}}\right)^2\xi_i\,
\Delta\left(\FLfull;\SFfull\right)
\nonumber \\*&&
\phantom{\lim_{\vec{k}_i\parallel\vec{k}_1}\left(1-y_i^2\right)}
+8\pi\as\mu^{2\ep}\left(\frac{2}{\sqrt{S}}\right)^2\xi_i
P_{S(a_1,\bar{a}_i)a_1}^{<}(1-\xi_i,\ep)
\nonumber \\*&&
\phantom{\lim_{\vec{k}_i\parallel\vec{k}_1}\left(1-y_i^2\right)}\times
\MTz\left(S(a_1,\bar{a}_i),a_2,\FLjbi;(1-\xi_i)k_1,k_2,\SFjbi\right).
\label{Mfourcolllim}
\eeqn
In this equation, the dependence of $\Delta$ upon the azimuthal variables
of parton $i$ is trivial; therefore, it can be shown that
\beq
\int d\Omega_i^{(2-2\ep)}\Delta=0.
\label{intodel}
\eeq
Furthermore, the second term in eq.~(\ref{Mfourcolllim}) does not depend
upon the azimuthal variables, and after the $d\Omega_i^{(2-2\ep)}$
integration in eq.~(\ref{sigcpm}) we get a factor
\beq
\int d\Omega_i^{(2-2\ep)}=\frac{2\pi^{1-\ep}}{\Gamma(1-\ep)}\,.
\label{intoom}
\eeq
Putting all the numerical factors together, taking into account
the formal limit
\beq
\lim_{\vec{k}_i\parallel\vec{k}_1}d\phi=
d\phi_3\left((1-\xi_i)k_1,k_2\to\SFjbi\right),
\eeq
using eq.~(\ref{S4clllim}) and
\beq
\frac{1}{\epb}=\frac{1}{\ep}-\gamma_E+\log 4\pi
\eeq
we get
\beqn
d\sigma_{\aoat,i}^{(in,+)}&=&-\frac{\as}{2\pi}\FLFsum\left(
\frac{1}{\epb}-\log\frac{S\delta_{\sss I}}{2\mu^2}\right)
\Bigg[\uoxiic-2\ep\uoxiilc\Bigg]
\xi_i P_{S(a_1,\bar{a}_i)a_1}^{<}(1-\xi_i,\ep)
\nonumber \\*&\times&
\uoffct
\MTz\left(S(a_1,\bar{a}_i),a_2,\FLjbi;(1-\xi_i)k_1,k_2,\SFjbi\right)
\nonumber \\*&\times&
{\cal S}_3 ([i]) d\phi_3\left((1-\xi_i)k_1,k_2\to\SFjbi\right)d\xi_i\,.
\label{sigcpole}
\eeqn
When the sum over $a_i$ is performed in eq.~(\ref{sigcpole}),
the quantity $S(a_1,\bar{a}_i)$ takes all the possible flavour
values such that $P_{S(a_1,\bar{a}_i)a_1}^{<}$ is different
from zero. Therefore, eq.~(\ref{sigcpole}) can be rewritten as
\beqn
d\sigma_{\aoat,i}^{(in,+)}&=&-\frac{\as}{2\pi}\sum_{d}\left(
\frac{1}{\epb}-\log\frac{S\delta_{\sss I}}{2\mu^2}\right)
\Bigg[\uoxiic-2\ep\uoxiilc\Bigg]
\xi_i P_{da_1}^{<}(1-\xi_i,\ep)
\nonumber \\*&\times&
\uoffct\FLFsumbi
\MTz\left(d,a_2,\FLjbi;(1-\xi_i)k_1,k_2,\SFjbi\right)
\nonumber \\*&\times&
{\cal S}_3 ([i]) d\phi_3\left((1-\xi_i)k_1,k_2\to\SFjbi\right)d\xi_i\,.
\label{sigcpole2}
\eeqn
We can now show that the divergent part of eq.~(\ref{sigcpole2})
is cancelled by the collinear counterterm of eq.~(\ref{cnt1}) up
to a factor $1/4$ (see eq.~(\ref{dsighatreal}) to understand the
$1/4$ factor). To this purpose,
we rewrite eq.~(\ref{cnt1}) in terms of the variable $\xi_i$.
Using eq.~(\ref{xieqz}), it is a simple matter of algebra to
prove the following identity
\beq
\delta(1-y_i)\left(\frac{1}{1-z}\right)_+ =
\delta(1-y_i)\Bigg[\uoxiic
+\delta(\xi_i)\log\xi_{cut}\Bigg]\,,
\label{uoxitouoz}
\eeq
where the $\delta(1-y_i)$ formally states the fact that the above
equation is strictly valid only in the collinear limit.
The Altarelli-Parisi kernel can be put in the form
\beqn
P_{ab}(z,0)&=&\frac{(1-z)P_{ab}^{<}(z,0)}{(1-z)_+}
+ \gamma(a)\delta_{ab}\delta(1-z)\,,
\label{AP1}
\\*
2C(a)\delta_{ab}\delta(1-z)&=&\delta(1-z)(1-z)P_{ab}^{<}(z,0)\,,
\label{AP2}
\eeqn
where the quantities $C(a)$ and $\gamma(a)$ were given in
eq.~(\ref{gCasimir}) for gluons and eq.~(\ref{qCasimir}) for quarks.
Using eq.~(\ref{uoxitouoz}), (\ref{AP1}) and (\ref{AP2}), and changing
the integration variable $z\to 1-\xi_i$, it is straightforward to
see that eq.~(\ref{cnt1}) can be cast in the form
\beqn
d\sigma_{\aoat}^{(cnt+)}&=&
\frac{\as}{2\pi}\sum_d\Bigg\{-K_{da_1}(1-\xi_i)
\nonumber \\*&&
+\frac{1}{\epb}\Bigg[\uoxiic \xi_i P_{da_1}^{<}(1-\xi_i,0)
+\delta_{da_1}\delta(\xi_i)\Big(\gamma(d)+2C(d)\log\xi_{cut}\Big)
\Bigg]\Bigg\}
\nonumber \\*&\times&
\uotfct\FLFsumbi
\MTz\left(d,a_2,\FLjbi;(1-\xi_i)k_1,k_2,\SFjbi\right)
\nonumber \\*&\times&
{\cal S}_3([i]) d\phi_3 \left((1-\xi_i)k_1,k_2\to\SFjbi\right)d\xi_i\,,
\label{cnt1xi}
\eeqn
since the dependence upon the index $i$ is immaterial.
Eq.~(\ref{cnt1xi}) is of the same form of eq.~(\ref{sigcpole2}).
It is now apparent that the quantity
\beq
d\hat\sigma_{\aoat,i}^{(in,+)}=d\sigma_{\aoat,i}^{(in,+)}
+\frac{1}{4}d\sigma_{\aoat}^{(cnt+)}
\label{siginplussplit}
\eeq
does not contain any purely collinear pole (i.e., proportional to the
Altarelli-Parisi kernel for $z<1$). Expanding in a Taylor series
eq.~(\ref{siginplussplit}), and using
\beq
P_{ab}^{<}(z,\ep)=P_{ab}^{<}(z,0)+\ep P_{ab}^{\prime <}(z,0)\,,
\eeq
we have
\beqn
d\hat\sigma_{\aoat,i}^{(in,+)}
&=&\frac{\as}{2\pi}\frac{1}{\epb}
\Big(\gamma(a_1)+2C(a_1)\log\xi_{cut}\Big)
\nonumber \\*&&\times
\uoffct\FLFsumbi\MTz\left(\FLfullbi;\SFfullbi\right)
{\cal S}_3([i]) d\phi_3 \left(k_1,k_2\to\SFjbi\right)
\nonumber \\*&+&
\frac{\as}{2\pi}\sum_{d}\Bigg\{\xi_i P_{da_1}^{<}(1-\xi_i,0)
\Bigg[\uoxiic\log\frac{S\delta_{\sss I}}{2\mu^2}
+2\uoxiilc\Bigg]
\nonumber \\*&&\phantom{\frac{\as}{2\pi}\sum_{d}}
-\xi_i P_{da_1}^{\prime <}(1-\xi_i,0)\uoxiic
-K_{da_1}(1-\xi_i)\Bigg\}
\nonumber \\*&&\times
\uoffct\FLFsumbi
\MTz\left(d,a_2,\FLjbi;(1-\xi_i)k_1,k_2,\SFjbi\right)
\nonumber \\*&&\times
{\cal S}_3 ([i]) d\phi_3\left((1-\xi_i)k_1,k_2\to\SFjbi\right) d\xi_i\,.
\label{sigchat}
\eeqn
The pole part of this equation, which originates from the $\delta$
term in the flavour-diagonal Altarelli-Parisi kernels, cancels a
corresponding term in the soft-virtual contribution. The remaining
part is finite, and can be numerically evaluated.
The dependence upon the index $i$ in the RHS of eq.~(\ref{sigchat})
is trivial, and the sum over $i$ is very easily carried out.
After some algebra, we get
\beqn
d\hat\sigma_{\aoat}^{(in,+)}&=&\sum_i d\hat\sigma_{\aoat,i}^{(in,+)}
\nonumber \\&=&
\frac{\as}{2\pi}\Bigg(\SVfact\,\frac{1}{\ep}-\log\frac{\mu^2}{Q^2}\Bigg)
\nonumber \\*&&\times
\Big(\gamma(a_1)+2C(a_1)\log\xi_{cut}\Big)
\FLTsum d\sigma^{(0)}(\FLTfull;\SFTfull)
\nonumber \\*&+&
\frac{\as}{2\pi}\sum_{d}\Bigg\{\xi P_{da_1}^{<}(1-\xi,0)
\Bigg[\uoxic\log\frac{S\delta_{\sss I}}{2\mu^2}
+2\uoxilc\Bigg]
\nonumber \\*&&\phantom{\frac{\as}{2\pi}\sum_{d}}
-\xi P_{da_1}^{\prime <}(1-\xi,0)\uoxic
-K_{da_1}(1-\xi)\Bigg\}
\nonumber \\*&&\times
\FLTsum d\sigma^{(0)}(d,a_2,\FLTj;(1-\xi)k_1,k_2,\STj)\,d\xi\,,
\label{collplusfinal}
\eeqn
where $d\sigma^{(0)}$ was defined in eq.~(\ref{bornfldef}).
In the very same way, we have
\beqn
d\hat\sigma_{\aoat}^{(in,-)}&=&
\frac{\as}{2\pi}\Bigg(\SVfact\,\frac{1}{\ep}-\log\frac{\mu^2}{Q^2}\Bigg)
\nonumber \\*&&\times
\Big(\gamma(a_2)+2C(a_2)\log\xi_{cut}\Big)
\FLTsum d\sigma^{(0)}(\FLTfull;\SFTfull)
\nonumber \\*&+&
\frac{\as}{2\pi}\sum_{d}\Bigg\{\xi P_{da_2}^{<}(1-\xi,0)
\Bigg[\uoxic\log\frac{S\delta_{\sss I}}{2\mu^2}
+2\uoxilc\Bigg]
\nonumber \\*&&\phantom{\frac{\as}{2\pi}\sum_{d}}
-\xi P_{da_2}^{\prime <}(1-\xi,0)\uoxic
-K_{da_2}(1-\xi)\Bigg\}
\nonumber \\*&&\times
\FLTsum d\sigma^{(0)}(a_1,d,\FLTj;k_1,(1-\xi)k_2,\STj)\,d\xi\,.
\label{collminusfinal}
\eeqn
\subsection{Final state singularities}

We finally turn to the problem of regulating the divergencies
appearing in eq.~(\ref{sigijout}). To this purpose,
we rewrite eq.~(\ref{kidef}) in the following way
\beq
k_i=\frac{\sqrt{S}}{2}\xi_i\left(1,\hat{k}_i\right),\;\;
\hat{k}_i=\hat{p}R\,,\;\;\hat{p}=\left(\vec{0},1\right)\,.
\eeq
Here $\hat{p}$ is a $3-2\ep$ dimensional vector and $R$ a $3-2\ep$
dimensional rotation matrix. With $\ep=0$ and eq.~(\ref{kidef}), the
explicit form of $R$ can be worked out, but we will not need it in the
following. We then parametrize the momentum of parton $j$ as
\beq
k_j=\frac{\sqrt{S}}{2}\xi_j\left(1,\hat{k}_j\right),\;\;
\hat{k}_j=\hat{p}_j R\,,\;\;
\hat{p}_j=\left(\sqrt{1-y_j^2}\vec{e}_{j{\sss T}},y_j\right).
\label{kjdef}
\eeq
{}From this definition we get
\beq
k_i\cdot k_j=\left(\frac{\sqrt{S}}{2}\right)^2\xi_i\xi_j\left(1-y_j\right).
\eeq
Therefore, the collinear limit $i\parallel j$ is obtained with the
parametrization of eq.~(\ref{kjdef}) by letting $y_j\to 1$. It
follows that we can regulate final state divergencies by multiplying
the invariant amplitude squared by the factor $(1-y_j)$. We also write
\beqn
d\phi&=&d\tilde{\phi} d\phi(j)
\\*&=&
d\tilde{\phi}\,\frac{1}{2(2\pi)^{3-2\ep}}
\left(\frac{\sqrt{S}}{2}\right)^{2-2\ep}\xi_j^{1-2\ep}
\left(1-y_j^2\right)^{-\ep} d\xi_j dy_j d\Omega_j^{(2-2\ep)}\,,
\eeqn
where, from eq.~(\ref{dphidef}),
\beq
d\tilde{\phi}=
(2\pi)^{4-2\ep}\delta^{4-2\ep}\left(k_1+k_2-\sum_{l=3}^6 k_l\right)
\prod_l^{[ij]}\frac{d^{3-2\ep}k_l}{(2\pi)^{3-2\ep}\,2k_l^0}\,.
\label{dphitildedef}
\eeq
Eq.~(\ref{sigijout}) becomes
\beqn
d\sigma_{\aoat,ij}^{(out)}&=&\uoffct\Bigg[\uoxiic-2\ep\uoxiilc\Bigg]
\left((1-y_j)\xi_i^2\FLFsum\MF(\FLfull)\right)
\nonumber \\*&\times&
{\cal S}_{ij}^{(1)} \Th(d_j-d_i) d\tilde{\phi}
\left(\frac{1}{2(2\pi)^{3-2\ep}}
\left(\frac{\sqrt{S}}{2}\right)^{2-2\ep}\right)^2
\left(1-y_i^2\right)^{-\ep} d\xi_i dy_i d\Omega_i^{(2-2\ep)}
\nonumber \\*&\times&
\xi_j^{1-2\ep}\left(1-y_j\right)^{-1-\ep}\left(1+y_j\right)^{-\ep}
d\xi_j dy_j d\Omega_j^{(2-2\ep)}\,.
\label{sigoutfull}
\eeqn
We can therefore use the identity
\beq
\left(1-y_j\right)^{-1-\ep}=-\frac{(\delta_o)^{-\ep}}{\ep}\delta(1-y_j)
+\uoyjmdo + {\cal O}(\ep),
\eeq
where $0<\delta_o\leq 2$, to split $d\sigma_{\aoat,ij}^{(out)}$
into two terms
\beq
d\sigma_{\aoat,ij}^{(out)}=d\sigma_{\aoat,ij}^{(out,+)}
+d\sigma_{\aoat,ij}^{(out,f)}\,,
\eeq
where $d\sigma_{\aoat,ij}^{(out,+)}$ is proportional to $\delta(1-y_j)$
and $d\sigma_{\aoat,ij}^{(out,f)}$ is free of singularities (and therefore
we can set $\ep=0$) and can be numerically integrated. Explicitly
\beqn
d\sigma_{\aoat,ij}^{(out,f)}&=&\uoffct\uoxiic\uoyjmdo
\left((1-y_j)\xi_i^2\xi_j\FLFsum\MF(\FLfull)\right)
{\cal S}_{ij}^{(1)}\Th(d_j-d_i)
\nonumber \\*&\times&
\left(\frac{1}{2(2\pi)^3}\left(\frac{\sqrt{S}}{2}\right)^2\right)^2
d\tilde{\phi}d\xi_i d\xi_j dy_i dy_j d\varphi_i d\varphi_j\,,
\label{sigoutijfin}
\eeqn
and
\beq
d\sigma_{\aoat,ij}^{(out,+)}=-\frac{(2\delta_o)^{-\ep}}{\ep}
\delta(1-y_j)dy_j\,{\cal D}{\cal A}\,d\mu\,,
\label{sigoutijp}
\eeq
where
\beqn
{\cal D}&=&\uoffct\Bigg[\uoxiic-2\ep\uoxiilc\Bigg]\,,
\label{Ddef}
\\
{\cal A}&=&\left((1-y_j)\xi_i^2\FLFsum\MF(\FLfull)\right)
{\cal S}_{ij}^{(1)}\Th(d_j-d_i)\,,
\\
d\mu&=&\left(\frac{1}{2(2\pi)^{3-2\ep}}
\left(\frac{\sqrt{S}}{2}\right)^{2-2\ep}\right)^2
d\tilde{\phi}\,\xi_j^{1-2\ep}\left(1-y_i^2\right)^{-\ep}
d\xi_i d\xi_j dy_i d\Omega_i^{(2-2\ep)} d\Omega_j^{(2-2\ep)}\,.
\nonumber \\*&&
\eeqn
We perform the following change of variables
\beq
\xi_i=(1-z)\xi_7,\,\;\;\xi_j=z\xi_7,\,\;\;\Rightarrow\;\;
d\xi_i d\xi_j=\xi_7 d\xi_7 dz\,.
\label{rhozdef}
\eeq
With this definition, we have
\beq
\delta(1-y_j)\left(k_i+k_j\right)\equiv \delta(1-y_j)k_7=\delta(1-y_j)
\frac{\sqrt{S}}{2}\xi_7\left(1,\sqrt{1-y_i^2}
\vec{e}_{i{\sss T}},y_i\right),
\label{kseven}
\eeq
which defines the momentum $k_7$; notice that the definition of
$\xi_7$ is therefore consistent with eq.~(\ref{kidef}); in the
collinear limit, this parameter is proportional to the
energy of the parton which eventually splits into the two collinear
partons $i$ and $j$. With the definition of $z$ in eq.~(\ref{rhozdef})
we have, analogously to eq.~(\ref{limcoll}),
\beqn
\lim_{y_j\to 1}\MF(\FLfull;\SFfull)&=&
\frac{4\pi\as\mu^{2\ep}}{k_i\cdot k_j}
\Delta\left(\FLfull;\SFfull\right)
\phantom{\MTz\left(\FLCollfull;\SCollfull\right)}
\nonumber \\*
&+&\frac{4\pi\as\mu^{2\ep}}{k_i\cdot k_j}
P_{a_j a_7}^{<}(z,\ep)\MTz\left(\FLCollfull;\SCollfull\right),
\label{limcollij}
\eeqn
with
\beq
a_7=S(a_i,a_j).
\eeq
Using eqs.~(\ref{limcollij}), (\ref{S4ijlimtmp})
and~(\ref{S4albelim}) we get
\beqn
\lim_{y_j\to 1}{\cal A}&=&
4\pi\as\mu^{2\ep}\left(\frac{\sqrt{S}}{2}\right)^{-2}
\frac{1-z}{z}\,\Th\left(z-\frac{1}{2}\right){\cal S}_3([ij])
\nonumber \\*&\times&
\FLFsum\Bigg[P_{a_j a_7}^{<}(z,\ep)
\MTz\left(\FLCollfull;\SCollfull\right)
+\Delta\left(\FLfull;\SFfull\right)\Bigg]\,,
\nonumber \\*&&
\eeqn
since from eq.~(\ref{rhozdef}) immediately follows
\beq
\Th(d_j-d_i)\equiv \Th\left(z-\frac{1}{2}\right).
\eeq
We can now exploit the following identities
\beqn
\uoxiic&=&\frac{1}{\xi_7}{\cal D}^{(0)}(z)\,,
\\*
\uoxiilc&=&\frac{1}{\xi_7}{\cal D}^{(1)}(z)\,,
\eeqn
where
\beqn
{\cal D}^{(0)}(z)&=&\uozm+
\log\left(\frac{\xi_7}{\xi_{cut}}\right)\delta(1-z),
\label{D0iden}
\\
{\cal D}^{(1)}(z)&=&\uozlm+\log\xi_7\uozm
+\frac{1}{2}\left(\log^2\xi_7-\log^2\xi_{cut}\right)\delta(1-z).
\nonumber \\*&&
\label{D1iden}
\eeqn
Finally, with the change of variables in eq.~(\ref{rhozdef}),
the measure becomes
\beq
d\mu=\left(\frac{1}{2(2\pi)^{3-2\ep}}
\left(\frac{\sqrt{S}}{2}\right)^{2-2\ep}\right)^2
d\tilde{\phi}\, z^{1-2\ep}dz\,\xi_7^{2-2\ep}d\xi_7
\left(1-y_i^2\right)^{-\ep} dy_i
d\Omega_i^{(2-2\ep)} d\Omega_j^{(2-2\ep)}\,.
\eeq
After some algebra, and using the analogous of
eqs.~(\ref{intodel}) and~(\ref{intoom})
when integrating over $d\Omega_j^{(2-2\ep)}$, we get
\beqn
d\sigma_{\aoat,ij}^{(out,+)}&=&-\frac{(2\delta_o)^{-\ep}}{\ep}
\frac{2\pi^{1-\ep}}{\Gamma(1-\ep)}
4\pi\as\mu^{2\ep}\left(\frac{\sqrt{S}}{2}\right)^{-2}
\uoffct\Bigg[{\cal D}^{(0)}(z)-2\ep\,{\cal D}^{(1)}(z)\Bigg]
\Th\left(z-\frac{1}{2}\right)
\nonumber \\*&\times&
dz\,(1-z)\,z^{-2\ep}\FLFsum P_{a_j a_7}^{<}(z,\ep)
\MTz\left(\FLCollfull;\SCollfull\right){\cal S}_3([ij])
\nonumber \\*&\times&
\left(\frac{1}{2(2\pi)^{3-2\ep}}
\left(\frac{\sqrt{S}}{2}\right)^{2-2\ep}\right)^2
d\tilde{\phi}\,\xi_7^{1-2\ep} d\xi_7 \left(1-y_i^2\right)^{-\ep}
dy_i d\Omega_i^{(2-2\ep)}\,.
\eeqn
Using eq.~(\ref{kseven}), we can see that the following relation holds
\beq
d\phi_3\left(k_1,k_2\to\SColl\right)=
\frac{1}{2(2\pi)^{3-2\ep}}\left(\frac{\sqrt{S}}{2}\right)^{2-2\ep}
\xi_7^{1-2\ep}\left(1-y_i^2\right)^{-\ep}
d\xi_7 dy_i d\Omega_i^{(2-2\ep)}\,d\tilde{\phi}\,.
\eeq
Therefore we have
\beqn
d\sigma_{\aoat,ij}^{(out,+)}&=&-\frac{\as}{2\pi}
\left(\frac{1}{\epb}-\log\frac{S\delta_o}{2\mu^2}\right)
\FLFsum\Bigg[{\cal I}_{a_j a_7}^{(0)}-2\ep\,{\cal I}_{a_j a_7}^{(1)}\Bigg]
\nonumber \\*&\times&
\uoffct\MTz\left(\FLCollfull;\SCollfull\right)
{\cal S}_3([ij]) d\phi_3\left(k_1,k_2\to\SColl\right),
\label{sigoutijp2}
\eeqn
where
\beqn
{\cal I}_{ab}^{(0)}&=&\int_{0}^{1} dz\, z^{-2\ep}\,(1-z)\,
P_{ab}^{<}(z,\ep)\,\Th\left(z-\frac{1}{2}\right){\cal D}^{(0)}(z)\,,
\label{I0int}
\\
{\cal I}_{ab}^{(1)}&=&\int_{0}^{1} dz\, z^{-2\ep}\,(1-z)\,
P_{ab}^{<}(z,\ep)\,\Th\left(z-\frac{1}{2}\right){\cal D}^{(1)}(z)\,.
\label{I1int}
\eeqn
Eq.~(\ref{sigoutijp2}) shows that the singular term due to final state
collinear emission factorizes into the product of two quantities;
a cross section, which has a two-to-three partonic kinematics, and a term
dependent upon the Altarelli-Parisi kernels, which completely describes
the collinear splitting. We can sum over $i,j$ the quantities
$d\sigma_{\aoat,ij}^{(out,+)}$. It is a simple matter of combinatorial
calculus to see that this sum can be rewritten in the following
way, by simply relabeling the partons
\beq
d\sigma_{\aoat}^{(out,+)}=
\sum_i\sum_j^{[i]}d\sigma_{\aoat,ij}^{(out,+)}=
\sum_{j=3}^{5}d\tilde{\sigma}_{\aoat,j}^{(out,+)}\,,
\eeq
where
\beqn
d\tilde{\sigma}_{\aoat,j}^{(out,+)}&=&-\frac{\as}{2\pi}
\left(\frac{1}{\epb}-\log\frac{S\delta_o}{2\mu^2}\right)
\FLTsum\sum_d\Bigg[{\cal I}_{d a_j}^{(0)}
-2\ep\,{\cal I}_{d a_j}^{(1)}\Bigg]
d\sigma^{(0)}(\FLTfull)\,.\phantom{aa}
\label{sigtildeoutj}
\eeqn
The expression for
$\sum_d [{\cal I}_{d a_j}^{(0)}-2\ep\,{\cal I}_{d a_j}^{(1)}]$
is given in appendix A. By substituting it in eq.~(\ref{sigtildeoutj})
we get
\beqn
d\sigma_{\aoat}^{(out,+)}&=&\frac{\as}{2\pi}\SVfact
\FLTsum d\sigma^{(0)}(\FLTfull)
\nonumber \\*&&\times
\frac{1}{\ep}\sum_{j=3}^{5}\Bigg[\gamma(a_j)
-2C(a_j)\log\frac{2E_j}{\xi_{cut}\sqrt{S}}\Bigg]
\nonumber \\*
&+&\frac{\as}{2\pi}\,\FLTsum d\sigma^{(0)}(\FLTfull)
\nonumber \\*&&\times
\sum_{j=3}^{5}\Bigg[\gamma^\prime(a_j)
-\log\frac{S\delta_o}{2Q^2}\left(\gamma(a_j)
-2C(a_j)\log\frac{2E_j}{\xi_{cut}\sqrt{S}}\right)
\nonumber \\*&&
+2C(a_j)\left(\log^2\frac{2E_j}{\sqrt{S}}-\log^2\xi_{cut}\right)
-2\gamma(a_j)\log\frac{2E_j}{\sqrt{S}}\Bigg].
\label{clloutfact}
\eeqn
The explicit expression for $\gamma^\prime(a_j)$
is given in eqs.~(\ref{gmmprimeglu}) and~(\ref{gmmprimeqrk}).
\section{Results}

In the previous sections we have shown that the subtracted
next-to-leading order contribution,
defined in eq.~(\ref{counterterms}), can be
written as the sum of a term with a two-to-four partonic
kinematics, and a term with a two-to-three partonic
kinematics
\beq
d\hat{\sigma}_{\aoat}^{(1)}=d\hat{\sigma}_{\aoat}^{(1,4p)}
+d\hat{\sigma}_{\aoat}^{(1,3p)}.
\label{finalsigfin}
\eeq
The two-to-four part is
\beq
d\hat{\sigma}_{\aoat}^{(1,4p)}=\sum_i\left(d\sigma_{\aoat,i}^{(fin)}
+d\sigma_{\aoat,i}^{(in,f)}
+\sum_j^{[i]} d\sigma_{\aoat,ij}^{(out,f)}\right),
\label{nlotwotofour}
\eeq
where the quantities appearing in the RHS of this equation
were given in eqs.~(\ref{sigrns}), (\ref{sigiinfin2})
and~(\ref{sigoutijfin}) respectively. The two-to-three part is
\beq
d\hat{\sigma}_{\aoat}^{(1,3p)}=
d\sigma_{\aoat}^{(v)}+d\sigma_{\aoat}^{(s)}
+d\hat{\sigma}_{\aoat}^{(in,+)}+d\hat{\sigma}_{\aoat}^{(in,-)}
+d\sigma_{\aoat}^{(out,+)}
\label{nlotwotothree}
\eeq
where the quantities appearing in the RHS of this equation
were given in eqs.~(\ref{virtjetdef}) (with the matrix
element of eq.~(\ref{virttransamp})), (\ref{softfact}),
(\ref{collplusfinal}), (\ref{collminusfinal}) and~(\ref{clloutfact}).
Every term in the RHS of eq.~(\ref{nlotwotofour}) is finite,
and can be numerically integrated. On the other hand, the
quantities in the RHS of eq.~(\ref{nlotwotothree}) are divergent;
nevertheless, from their explicit expression previously reported,
it is apparent that the divergencies cancel in the sum, and
$d\hat{\sigma}_{\aoat}^{(1,3p)}$ is finite. The
finite part of eqs.~(\ref{collplusfinal}) and~(\ref{collminusfinal}),
due to initial state collinear contribution, is slightly more
complicated than the usual two-to-three kinematics contribution,
since it has an additional folding in the variable $\xi$.
Therefore, we further split eq.~(\ref{nlotwotothree})
into a sum of two terms, one of which does not contain any folding.
The reason for doing so is that in a numerical computation
the two terms have to be differently treated, in spite of the
fact of having the same final state kinematics. We have
\beq
d\hat{\sigma}_{\aoat}^{(1,3p)}=d\hat{\sigma}_{\aoat}^{(1,3pv)}
+d\hat{\sigma}_{\aoat}^{(1,3pr)}.
\eeq
The part without $\xi$ folding is
\beqn
d\hat{\sigma}_{\aoat}^{(1,3pv)}&=&
\frac{\as}{2\pi}\FLTsum {\cal Q}(\FLTfull) d\sigma^{(0)}(\FLTfull)
\nonumber \\*&+&
\frac{\as}{2\pi}\,\frac{1}{2}\sum_{\stackrel{n,m=1}{n\neq m}}^{5}
{\cal I}_{mn}^{(reg)}\FLTsum d\sigma_{mn}^{(0)}(\FLTfull)
\nonumber \\*&+&
\frac{\as}{2\pi}\uotfct\FLTsum
\MTo_{\sss NS}(\FLTfull)\,{\cal S}_3\,d\phi_3\,,
\eeqn
where
\beqn
{\cal Q}(\FLTfull)&=&\sum_{j=3}^{5}\Bigg[\gamma^\prime(a_j)
-\log\frac{S\delta_o}{2Q^2}\left(\gamma(a_j)
-2C(a_j)\log\frac{2E_j}{\xi_{cut}\sqrt{S}}\right)
\nonumber \\*&&
+2C(a_j)\left(\log^2\frac{2E_j}{\sqrt{S}}-\log^2\xi_{cut}\right)
-2\gamma(a_j)\log\frac{2E_j}{\sqrt{S}}\Bigg]
\nonumber \\*&-&
\log\frac{\mu^2}{Q^2}
\Bigg(\gamma(a_1)+2C(a_1)\log\xi_{cut}
+\gamma(a_2)+2C(a_2)\log\xi_{cut}\Bigg),\phantom{aa}
\eeqn
and $d\sigma^{(0)}$, $d\sigma_{mn}^{(0)}$ were defined in
eqs.~(\ref{bornfldef}) and~(\ref{bornmndef}) respectively.
The part with the folding can directly be read from
eqs.~(\ref{collplusfinal}) and~(\ref{collminusfinal}),
and it is
\beqn
d\hat{\sigma}_{\aoat}^{(1,3pr)}&=&
\frac{\as}{2\pi}\sum_{d}\Bigg\{\xi P_{da_1}^{<}(1-\xi,0)
\Bigg[\uoxic\log\frac{S\delta_{\sss I}}{2\mu^2}
+2\uoxilc\Bigg]
\nonumber \\*&&\phantom{\frac{\as}{2\pi}\sum_{d}}
-\xi P_{da_1}^{\prime <}(1-\xi,0)\uoxic
-K_{da_1}(1-\xi)\Bigg\}
\nonumber \\*&&\times
\FLTsum d\sigma^{(0)}(d,a_2,\FLTj;(1-\xi)k_1,k_2,\STj)\,d\xi
\nonumber \\*&+&
\frac{\as}{2\pi}\sum_{d}\Bigg\{\xi P_{da_2}^{<}(1-\xi,0)
\Bigg[\uoxic\log\frac{S\delta_{\sss I}}{2\mu^2}
+2\uoxilc\Bigg]
\nonumber \\*&&\phantom{\frac{\as}{2\pi}\sum_{d}}
-\xi P_{da_2}^{\prime <}(1-\xi,0)\uoxic
-K_{da_2}(1-\xi)\Bigg\}
\nonumber \\*&&\times
\FLTsum d\sigma^{(0)}(a_1,d,\FLTj;k_1,(1-\xi)k_2,\STj)\,d\xi\,.
\eeqn

We have derived our results in the partonic center-of-mass frame.
This slightly constrains the general validity of the formalism, since
in the numerical computations $x_1$ and $x_2$, the Bjorken $x$ of the
incoming partons, have to be chosen as independent integration variables.
Notice that they define the boost from the hadronic center-of-mass
frame to the partonic one.
We stress that this constraint can be very easily relaxed, without
affecting the correctness of the derivation. Nevertheless, as far
as the numerical computations are concerned, it appears to be
advantageous to work in the partonic center-of-mass frame.
We remind that in previous applications of the subtraction method
to jet physics $x_1$ and $x_2$ have been expressed in terms
of transverse momenta and rapidities.
With our finite next-to-leading order partonic cross sections,
we can get the physical cross section using eq.~(\ref{factth}).

A final remark on the numerical computation: the parameters
$\xi_{cut}$, $\delta_{\sss I}$ and $\delta_{\sss O}$ define
the soft and collinear subtractions.
Thanks to this property, we can freely redefine the terms
in the RHS of eq.~(\ref{finalsigfin}) without affecting the sum.
This gives us the possibility of a significant numerical check
on the correctness of the calculation; the physical results
have to be independent from the value of these parameters.
Also, notice that a clever choice of the parameters results
in saving computing time. If for example eq.~(\ref{uoxidef})
is considered, it is apparent that in most cases the soft counterterm
($f(0)$) is not evaluated if $\xi_{cut}$ is small. On the other
hand, $\xi_{cut}$ should not have a value too close to $0$, because
in this way the quality of the convergence is rather poor.
\section{Conclusions}

We have studied the production of three-jet inclusive quantities
at the next-to-leading order in QCD, using the subtraction method.
Comparing to previous treatments~[\ref{kssub}], in which the one- or
two-jet production is dealt with this method, we used a
somewhat different approach in at least two important aspects.
First of all, we used angle and energy variables, instead of
transverse momenta and rapidities; this results in a
simplification when integrating the eikonal factors, which
appear whenever one of the final state partons gets soft.
As a second feature, we fully exploited the measurement function,
which defines infrared-safe observables, to disentangle the soft
and collinear singularities in the real contribution. In fact, the
measurement function can always be cast as a sum of terms,
each of them getting contributions only from the infrared
singular regions associated with a given parton. At the very end,
we have therefore splitted the cross section into single-singular
contributions, without the drawback of decomposing the transition
amplitude squared with partial fractioning.

The analytical results obtained were organized in a form
suited for numerical computations. We splitted the real
plus virtual contribution in a sum of terms with different final
state partonic kinematics. The possibility has been left to redefine
them up to a finite piece, without affecting the sum, which is the
only quantity being physically meaningful.

\enlargethispage*{1000pt}
Although we treated explicitly only three-jet production in
hadron-hadron collisions, our formalism can be
easily extended to $n$-jet production. Furthermore, we emphasize
that it is also applicable to $n$-jet production in
$e^+e^-$ annihilation and in photon-hadron collisions.
\newpage
\appendix
\section{Soft and collinear integrals}

In this appendix, we collect the results for the integrals
of the eikonal factors, which are substituted
into eq.~(\ref{sigsoft2}), and for the integrals of the
Altarelli-Parisi kernels, with the prescription defined
in eq.~(\ref{I0int}) and (\ref{I1int}).

$\bullet$~{\it Soft integrals}: given eq.~(\ref{sigsoft2}),
we will consider the quantities
\beq
{\cal J}_{nm}=\int E_{nm} d\omega_i\,,
\eeq
where
\beqn
E_{nm}&=&\left(\frac{\sqrt{S}}{2}\right)^2
\frac{k_n\cdot k_m}{k_n\cdot k_i~k_m\cdot k_i}\,\xi_i^2\,,
\\*
d\omega_i&=&\left(1-y_i^2\right)^{-\ep} dy_i d\Omega_i^{(2-2\ep)}\,.
\eeqn
Using eq.~(\ref{kidef}) we can rewrite
\beq
E_{nm}=\frac{1-\cos\theta_{nm}}
{\left(1-\cos\theta_{ni}\right)\left(1-\cos\theta_{mi}\right)}\,,
\label{Iconalnm}
\eeq
where $\theta_{\alpha\beta}$ is the angle between the directions
of the three-momenta $\vec{k}_\alpha$ and $\vec{k}_\beta$.
The measure $d\omega_i$ is the angular measure in $3-2\ep$
dimensions for the momentum $k_i$; since also the eikonal factor
is rotationally invariant, we can redefine at will the
angular variables of the momentum $\vec{k}_i$. If we choose
\beq
\vec{k}_i=\left(..,\sin\varphi\sin\theta,\cos\varphi\sin\theta,
\cos\theta\right)\,,
\eeq
then we have
\beq
d\omega_i=2^{1-2\ep}\pi^{-\ep}\frac{\Gamma(1-\ep)}{\Gamma(1-2\ep)}
\sin^{-2\ep}\varphi\sin^{-2\ep}\theta\,d\cos\theta\,d\varphi\,,
\eeq
where an integration has been carried out over the variables
upon which the eikonal factor can not depend. Finally, we
decompose eq.~(\ref{Iconalnm}) as follows
\beq
E_{nm}=\frac{1}{1-\cos\theta_{ni}}+\frac{1}{2}\left(E_{nm}
-\frac{1}{1-\cos\theta_{ni}}-\frac{1}{1-\cos\theta_{mi}}
\right)\,+\,(n\,\leftrightarrow\,m)\,.
\eeq
The first term in the RHS of this equation will give, after integration,
a collinear pole. The second term is regulated by means of the
subtractions, and will contribute a finite term to the integral.
The final result is
\beqn
{\cal J}_{nm}&=&2^{1-2\ep}\pi^{-\ep}\frac{\Gamma(1-\ep)}{\Gamma(1-2\ep)}
\nonumber \\*&\times&
\Bigg[\frac{2^{2\ep}\pi\Gamma(1-2\ep)}{\left(\Gamma(1-\ep)\right)^2}
\left(-\frac{2}{\ep}+4\log 2
+2\ep\left(\frac{\pi^2}{6}-2\log^2 2\right)\right)
+2\pi\log\left(\frac{1-\cos\theta_{nm}}{2}\right)
\nonumber \\*&&
-2\ep\,\pi\left(-{\rm Li}_2\left(\frac{1-\cos\theta_{nm}}{2}\right)
+\frac{1}{2}\log^2\Bigg(2\left(1-\cos\theta_{nm}\right)\Bigg)
+\frac{\pi^2}{6}-2\log^2 2\right)
\nonumber \\*&&
+2\ep\,\pi\log\Bigg(2\left(1+\cos\theta_{nm}\right)\Bigg)
\log\left(\frac{1-\cos\theta_{nm}}{2}\right)\Bigg].
\label{Icnlnm}
\eeqn
It is just a matter of trivial algebra to get from
this equation the expressions of eqs.~(\ref{Imndiv})
and~(\ref{Imnreg}).

$\bullet$~{\it Collinear integrals}: the calculation of the integrals
defined in eqs.~(\ref{I0int}) and~(\ref{I1int}) is a lengthy but
trivial operation. Using eqs.~(\ref{D0iden}) and~(\ref{D1iden}),
and taking into account that the
relabeling of the partons amounts to the substitution
$\xi_7\to\xi_j$ in the expressions of ${\cal I}_{d a_j}^{(0)}$ and
${\cal I}_{d a_j}^{(1)}$, and that by construction
\beq
\xi_j=\frac{2E_j}{\sqrt{S}}\,,
\eeq
we get
\beq
{\cal Z}(a_j)=\sum_d\Bigg[{\cal I}_{d a_j}^{(0)}
-2\ep\,{\cal I}_{d a_j}^{(1)}\Bigg]
\eeq
where
\beqn
{\cal Z}(a_j)&=&2C(a_j)\log\frac{2E_j}{\xi_{cut}\sqrt{S}}-\gamma(a_j)
\nonumber \\*&-&
\ep\left[\gamma^\prime(a_j)-2\gamma(a_j)\log\frac{2E_j}{\sqrt{S}}
+2C(a_j)\left(\log^2\frac{2E_j}{\sqrt{S}}-\log^2\xi_{cut}\right)\right],
\phantom{aaa}
\eeqn
with
\beqn
\gamma^\prime(g)&=&\frac{67}{9}\CA-\frac{2\pi^2}{3}\CA
-\frac{23}{9}\TF N_{f}\,,
\label{gmmprimeglu}
\\*
\gamma^\prime(q)&=&\frac{13}{2}\CF-\frac{2\pi^2}{3}\CF\,.
\label{gmmprimeqrk}
\eeqn
\newpage
\section{Collinear limits}

In section 4 we have stated that the $\Delta$ term of
eqs.~(\ref{limcoll}) and (\ref{limcollij}) does not
contribute to the result, since its average over the
azimuthal angle of the collinearly emitted parton is zero.
This fact can be proved in $4-2\ep$ dimensions studying the
collinear limit in terms of the Sudakov variables (see e.g.
ref.~[\ref{kssub}]). In spite of the fact that the contribution
of $\Delta$ is zero, it is of some importance to know its explicit
form, since it appears in local subtraction terms where it
improves the numerical treatment of equations like,
for example, eq.~(\ref{sigiinfin2}).
The goal of this appendix is to express $\Delta$ as a function
of the helicity amplitudes known from the literature.
We will work in $4$ dimensions, since we will eventually
use $\Delta$ only in numerical computations.

In the following, we will consider the production process
of $n$ particles
\beq
1\,+\,2\;\rightarrow\; X\,+\,i\,+\,j
\eeq
in the limit in which the two massless particles $i$ and $j$
become collinear to each other or collinear to the incoming
massless particles $1$ or $2$. Notice that the properties of
the $n-2$ particles denoted collectively by $X$ are
unspecified and of no importance in what follows. This implies that
the results of this section are completely general and their validity
is not restricted to jet physics.

We begin by considering the final state emission, that is
the case in which $i$ and $j$ are collinear. Therefore, we study
the structure of the process
\beq
P\;\rightarrow\; i\,j\,.
\eeq
We have to consider only three cases:
\beqn
{\rm a)}\;\;\;\; g\;\rightarrow\;g\,g\,,
\\*
{\rm b)}\;\;\;\; g\;\rightarrow\;q\,\bar{q}\,,
\\*
{\rm c)}\;\;\;\; q\;\rightarrow\;q\,g\,,
\eeqn
where in case c) $q$ can be either a quark or an antiquark.
We begin with case a). Giving color indices
$b$ and $c$ to the splitted gluons $i$ and $j$, and defining
the fraction of momentum $z$ through the equations
\beq
k_i=zk_{\sss P}\,,\;\;\;\;k_j=(1-z)k_{\sss P}\,,
\label{zdef}
\eeq
the $n$-particle transition amplitude can be written as
\beq
\An(h_i,h_j,\Hnd)\LC\gs\Sumae\Sumhe C(d_e,b,c)
\Pgghhh(z)\Anu_{d_{e}}(h_e,\Hnd)\,.
\label{ampn}
\eeq
Although the structure of this equation remains valid at all
orders in perturbation theory, we will restrict in the following
to the leading-order splitting. Therefore,
$C(d_e,b,c)$ is the color factor of the $ggg$ vertex, $\Pgghhh$
is the leading-order splitting function for fixed helicities
(which can be obtained from the entries of
table~\ref{splitting_tab} as explained in the caption)
and $\Anu_{d_{e}}$ is the transition amplitude for the
process $1+2\to X+g$, when the emitted gluon has color $d_e$.
The dependence upon the color indices $\{d_l\}$ of the remaining
$n-2$ particles is not explicitly indicated in $\Anu$.
In the following, we will systematically suppress the indication of
the dependence upon all color labels with the exception of $d_e$,
which is the only one relevant for our purposes.
The quantity $\Hnd$ is the set of the helicities of the $n-2$
particles which are not involved in the splitting process.
Notice that we only consider the range $z<1$, where the splitting
functions with fixed helicities are meaningful.
\begin{table}
\begin{center}
\begin{tabular}{|l||c|c|c|} \hline
$(h,h_a,h_b)$ & $g\,\to\,gg$ & $g\,\to\,q\bar{q}$ & $q\,\to\,qg$
\\ \hline \hline
$(+,+,+)$ & $[ab]$
          & $0$
          & $z^{1/2}\,[ab]$
\\ \hline
$(+,+,-)$ & $-z^2\,<ab>$
          & $z^{1/2}\,(1-z)^{3/2}\,<ab>$
          & $-z^{3/2}\,<ab>$
\\ \hline
$(+,-,+)$ & $-(1-z)^2\,<ab>$
          & $-z^{3/2}\,(1-z)^{1/2}\,<ab>$
          & $0$
\\ \hline
$(+,-,-)$ & $0$
          & $0$
          & $0$
\\ \hline
$(-,+,+)$ & $0$
          & $0$
          & $0$
\\ \hline
$(-,+,-)$ & $(1-z)^2\,[ab]$
          & $z^{3/2}\,(1-z)^{1/2}\,[ab]$
          & $0$
\\ \hline
$(-,-,+)$ & $z^2\,[ab]$
          & $-z^{1/2}\,(1-z)^{3/2}\,[ab]$
          & $z^{3/2}\,[ab]$
\\ \hline
$(-,-,-)$ & $-<ab>$
          & $0$
          & $-z^{1/2}\,<ab>$
\\ \hline
\end{tabular}
\ccaption{}{\label{splitting_tab}
Splitting process $P(h)\to a(h_a)b(h_b)$ for all the possible choices
of the partons $a$ and $b$ and the helicities $h$, $h_a$ and $h_b$. The
splitting functions $S_{\sss aP}^{h h_a h_b}$ are obtained by dividing
the entries of the table by $<ab>[ab]\sqrt{z(1-z)}$. By construction,
the parton $a$ has the fraction $z$ of the momentum of the
parton $P$. The splitting functions $S_{gq}$ can be obtained from
$S_{qq}$ with the formal substitution $z\to 1-z$.
}
\end{center}
\end{table}
Squaring eq.~(\ref{ampn}) and summing over the color of the
emitted gluon we get
\beqn
\abs{\An(h_i,h_j,\Hnd)}^2&\LC&\gs^2\sum_{b,b^\prime}\sum_{c,c^\prime}
\delta_{b b^\prime}\delta_{c c^\prime}
\Sumaeae\Sumhehe C(d_e,b,c) C^*(d_e^\prime,b^\prime,c^\prime)
\nonumber \\*&\times&
\Pgghhh(z)\left(\Pgghphh(z)\right)^*
\nonumber \\*&\times&
\Anu_{d_{e}}(h_e,\Hnd)
\left(\Anu_{d_{e}^\prime}(h_e^\prime,\Hnd)\right)^*\,.
\label{ampnsq}
\eeqn
Since $C(a,b,c)=f_{abc}$, and using the normalization conventions
of Mangano and Parke~[\ref{ManganoParkepr}], we have
\beq
\sum_{b,b^\prime}\sum_{c,c^\prime}
\delta_{b b^\prime}\delta_{c c^\prime}
C(d_e,b,c) C^*(d_e^\prime,b^\prime,c^\prime)=
2\CA\delta_{d_e d_e^\prime}\,.
\label{colorfactgg}
\eeq
Therefore, eq.~(\ref{ampnsq}) becomes
\beqn
\abs{\An(h_i,h_j,\Hnd)}^2&\LC&2\gs^2\CA\Sumae
\left(\Sumhe\Pgghhh(z)\Anu_{d_e}(h_e,\Hnd)\right)
\phantom{aaaa}
\nonumber \\*&&\phantom{2\CA}\times
\left(\Sumhep\Pgghphh(z)\Anu_{d_{e}}
(h_e^\prime,\Hnd)\right)^* .
\label{Antemp}
\eeqn
Writing explicitly the sums over $h_e$ and $h_e^\prime$
and carrying out the multiplication we obtain
\beq
\abs{\An(h_i,h_j,\Hnd)}^2\LC\abs{\Nn(h_i,h_j,\Hnd)}^2
+{\cal R}(h_i,h_j,\Hnd)\,,
\eeq
where
\beqn
\abs{\Nn(h_i,h_j,\Hnd)}^2&=&2\gs^2\CA\abs{\Pggplus(z)}^2
\Sumae\abs{\Anu_{d_e}(+,\Hnd)}^2
\nonumber \\*
&+&2\gs^2\CA\abs{\Pggminus(z)}^2\Sumae\abs{\Anu_{d_e}(-,\Hnd)}^2\,,
\label{Nndef}
\\
{\cal R}(h_i,h_j,\Hnd)&=&4\gs^2\CA\Re\Bigg\{\Pggplus(z)
\left(\Pggminus(z)\right)^*
\phantom{\left(\Anu_{d_e}(-,\Hnd)\right)^*}
\nonumber \\*&\times&
\Sumae \Anu_{d_e}(+,\Hnd)
\left(\Anu_{d_e}(-,\Hnd)\right)^*\Bigg\}\,.
\label{Deltadef}
\eeqn
By construction, given eq.~(\ref{ampn}), the quantity
\beq
\abs{\Anu(\pm,\Hnd)}^2=\sum_{\Icol}\Sumae\abs{\Anu_{d_e}(\pm,\Hnd)}^2
\eeq
is the $(n-1)$-particle amplitude squared and summed over colors,
for a given helicity configuration $(\pm,\Hnd)$. Therefore,
taking into account
\beq
\sum_{h_i,h_j}\abs{\Pggplus(z)}^2=\sum_{h_i,h_j}\abs{\Pggminus(z)}^2=
\frac{1}{\CA}\frac{1}{2k_i\cdot k_j}P_{gg}^<(z)\,,
\eeq
where $P_{gg}^<$ is the unpolarized Altarelli-Parisi kernel
for $z<1$, from eq.~(\ref{Nndef}) we get
\beq
\sum_{h_i,h_j,\{h_l\}}\sum_{\Icol}\abs{\Nn(h_i,h_j,\Hnd)}^2=
\frac{4\pi\as}{k_i\cdot k_j}P_{gg}^<(z)\abs{\Anu}^2\,,
\label{Nfinal}
\eeq
where
\beq
\abs{\Anu}^2=\sum_{h,\{h_l\}}\abs{\Anu(h,\Hnd)}^2
\eeq
is the full $(n-1)$-particle transition amplitude squared and
summed over helicities and colors. From it, we define the Born amplitude
squared by multiplying by the flux factor and averaging over the spin
and color of the incoming particles
\beq
\Mnu=\frac{1}{2\,k_1\cdot k_2}\,
\frac{1}{\omega(a_1)\,\omega(a_2)}\abs{\Anu}^2.
\label{Mamp}
\eeq
When considering three-jet production at the next-to-leading order,
this quantity coincides with $\MTz$, defined in eq.~(\ref{bornampdef}).
Eq.~(\ref{Deltadef}) can be simplified using the explicit
form of the splitting functions with fixed helicities. We have
\beq
\Pggplus(z)\left(\Pggminus(z)\right)^*=
\delta_{h_i\bar{h}_j}\,\frac{z(1-z)}{[ij]^2}\,.
\eeq
Therefore we get
\beqn
&&\sum_{h_i,h_j,\{h_l\}}\sum_{\Icol}{\cal R}(h_i,h_j,\Hnd)=
32\pi\as\CA\, z(1-z)
\phantom{\left(\Anu_{d_e}(-,\Hnd)\right)^* aaaaaaaa}
\nonumber \\*&&\phantom{aaaaaa}\times
\Re\Bigg\{\frac{1}{[ij]^2}\sum_{\{h_l\}}
\Sumaecl\Anu_{d_e}(+,\Hnd)
\left(\Anu_{d_e}(-,\Hnd)\right)^*\Bigg\}.
\label{Deltasum}
\eeqn
We can now write the full $n$-particle amplitude squared
\beq
\Mn=\frac{1}{2\,k_1\cdot k_2}\,
\frac{1}{\omega(a_1)\,\omega(a_2)}\abs{\An}^2.
\label{Mnplusamp}
\eeq
For three-jet production at the next-to-leading order, $\Mn$
coincides with $\MF$, defined in eq.~(\ref{realampdef}).
Using eqs.~(\ref{Nfinal}), (\ref{Mamp}) and (\ref{Deltasum})
we get
\beq
\Mn\LC\frac{4\pi\as}{k_i\cdot k_j}P_{gg}^<(z)\Mnu
-\frac{16\pi\as}{k_i\cdot k_j}\,\CA\,z(1-z)\tilde{{\cal M}}^{(n-1)}\,,
\label{fullcollgg}
\eeq
where
\beqn
&&\tilde{{\cal M}}^{(n-1)}=
\frac{1}{2\,k_1\cdot k_2}\,\frac{1}{\omega(a_1)\,\omega(a_2)}
\phantom{\Anu_{d_e}(+,\Hnd)
\left(\Anu_{d_e}(-,\Hnd)\right)^* aaaaaaaa}
\nonumber \\*&&\phantom{aaaaaa}\times
\Re\Bigg\{\frac{<ij>}{[ij]}\sum_{\{h_l\}}
\Sumaecl \Anu_{d_e}(+,\Hnd)
\left(\Anu_{d_e}(-,\Hnd)\right)^*\Bigg\}.
\label{Mtilde}
\eeqn
Notice that, apart from the factor $<ij>/[ij]$,
$\tilde{{\cal M}}^{(n-1)}$ would be equal to the Born amplitude
squared, eq.~(\ref{Mamp}), if the first entry of $\Anu$
were equal to the first entry of $(\Anu)^*$. In
eqs.~(\ref{Mamp}), (\ref{Mnplusamp})-(\ref{Mtilde}) we suppressed
the obvious functional dependence, but we emphasize that,
while ${\cal M}^{(n-1)}$ is independent from $k_i$ and $k_j$,
$\tilde{{\cal M}}^{(n-1)}$ has a residual dependence upon
these momenta through the factor \mbox{$<ij>/[ij]$}, which explicitly
appears in eq.~(\ref{Mtilde}).

We now turn to case b), that is the splitting $g\to q\bar{q}$.
We use again eq.~(\ref{zdef}) to define the $z$ parameter,
and assign color indices $k$ and $\bar{k}$ to the quark $i$
and to the antiquark $j$ respectively. Eq.~(\ref{ampn})
still holds, with the formal substitution
\beq
C(d_e,b,c)\,\rightarrow\, C(d_e,k,\bar{k})\,\equiv\,
\left(\lambda^{d_{e}}\right)_{k\bar{k}}\,.
\eeq
Obviously, we have also to substitute everywhere $P_{gg}^<$
with $P_{qg}^<$. Instead of eq.~(\ref{colorfactgg}) we have
\beq
\sum_{k,k^\prime}\sum_{\bar{k},\bar{k}^\prime}
\delta_{k k^\prime}\delta_{\bar{k} \bar{k}^\prime}
C(d_e,k,\bar{k}) C^*(d_e^\prime,k^\prime,\bar{k}^\prime)=
2\TF\delta_{d_e d_e^\prime}\,.
\label{colorfactqq}
\eeq
The derivation goes unchanged from this point on. We have only
to take into account that
\beq
\sum_{h_i,h_j}\abs{\Pqgplus(z)}^2=\sum_{h_i,h_j}\abs{\Pqgminus(z)}^2=
\frac{1}{\TF}\frac{1}{2k_i\cdot k_j}P_{qg}^<(z)\,,
\eeq
and
\beq
\Pqgplus(z)\left(\Pqgminus(z)\right)^*=
-\delta_{h_i\bar{h}_j}\,\frac{z(1-z)}{[ij]^2}\,.
\eeq
Eq.~(\ref{fullcollgg}) gets therefore modified as follows
\beq
\Mn\LC\frac{4\pi\as}{k_i\cdot k_j}P_{qg}^<(z)\Mnu
+\frac{16\pi\as}{k_i\cdot k_j}\,\TF\,z(1-z)\tilde{{\cal M}}^{(n-1)}\,.
\label{fullcollqg}
\eeq

Finally, we deal with the splitting $q\to qg$. With obvious
modifications due to the fact that the exchanged particle
in now a quark instead of a gluon, we can follow the derivation
outlined before. Nevertheless, we immediately understand
that in this case there is no ${\cal R}$ term. In fact, we know
from eq.~(\ref{Antemp}) that such a term is present if and only
if for a given external helicity configuration the exchanged virtual
particle can have both helicities. In the present case, this is forbidden
due to helicity conservation along the quark line. More formally,
for the $q\to qg$ splitting we have always
\beq
\Pgqplus(z)\left(\Pgqminus(z)\right)^*=
\Pqqplus(z)\left(\Pqqminus(z)\right)^*=0
\eeq
for every possible choice of $h_i$ and $h_j$.

We collect now the results obtained so far, also explicitly
indicating the dependence upon the momenta of the particles, referring
to eq.~(\ref{zdef}) for the definition of the kinematics:
\beqn
&&\Mn\left(k_1,k_2;\,..,k_i,..,k_j,..\right)\LC
\nonumber \\*&&\phantom{aaaaaaa+}
\frac{4\pi\as}{k_i\cdot k_j}\,P_{a_i S(a_i,a_j)}^<(z)
\Mnu\left(k_1,k_2;\,..,k_{\sss P},..\right)
\nonumber \\*&&\phantom{aaaaaaa}
+\frac{4\pi\as}{k_i\cdot k_j}\,Q_{a_i S(a_i,a_j)^\star}(z)
\tilde{{\cal M}}^{(n-1)}\left(k_1,k_2;\,..,k_i,..,k_j,..\right),
\label{resume}
\eeqn
where
\beqn
Q_{gg^\star}(z)&=&-4\CA\,z(1-z)\,,
\label{Q1}
\\
Q_{qg^\star}(z)&=&4\TF\,z(1-z)\,,
\\
Q_{gq^\star}(z)&=&0\,,
\\
Q_{qq^\star}(z)&=&0\,.
\label{Q4}
\eeqn
In these equations, the $^\star$ symbol over the flavour of the
particle that eventually splits reminds that this particle
is off-shell. In principle, this notation should be extended
also to the Altarelli-Parisi splitting kernels, but at the leading
order $P_{ab^\star}=P_{a^\star b}$, and therefore there is no
need to keep track of the off-shell particle. By construction,
the $\Delta$ term of section 4 is equal to $Q\,\tilde{{\cal M}}^{(n-1)}$.

We have now to deal with the collinear emission from an incoming
particle. We have two options: we can perform an explicit calculation,
as we have done for the final state emission; or otherwise we can exploit
the crossing symmetry property of the $n$-particle amplitude~[\ref{kssub}].
We will pursue this second option, and to be definite we will
restrict to emission from particle $1$. To proceed further, we have to
think in terms of an (unphysical) amplitude, for which also the
particle $1$ is outgoing; the limit $1\parallel j$ will therefore
be given by eq.~(\ref{resume}) (with obvious modifications in the
notation). The kinematics will be given by eq.~(\ref{zdef})
\beq
\bar{k}_1=zk_{\sss P}\,,\;\;\;\;k_j=(1-z)k_{\sss P}\,,
\eeq
where $\bar{k}_1=-k_1$ since in the unphysical amplitude
the particle $1$ has negative energy. In doing the
crossing, apart from the substitution $\bar{k}_1\to k_1$,
we will have to refer the kinematics to $k_1$, that is
\beq
k_j=(1-z)k_1\,,\;\;\;\;k_{\sss P}=zk_1
\label{zdefcr}
\eeq
or
\beq
z\,\longrightarrow\,1/z\,.
\eeq
Since $k_{\sss P}$ is the four-momentum of an incoming particle
in the $(n-1)$-particle amplitude, the second equation
in~(\ref{zdefcr}) implies that the flux factor of the $(n-1)$-particle
amplitude will be $1/z$ times the flux factor of the
$n$-particle amplitude. Therefore, to have the correct normalization,
we have to multiply by $z$ the $(n-1)$-particle amplitude.
Furthermore, since the flavour of $1$ and $P$ can be different,
a factor $\omega(a_{\sss P})/\omega(a_1)$ will take into account
the different color and spin normalization factors of the
$n$-particle and $(n-1)$-particle amplitudes. Eq.~(\ref{resume})
becomes therefore
\beqn
&&\Mn\left(k_1,k_2;\,..,k_j,..\right)\LCu
\nonumber \\*&&\phantom{aa}
-(-)^{[\sigma(a_1)+\sigma(S(a_1,\bar{a}_j)]}\,\frac{4\pi\as}{k_1\cdot k_j}\,
\frac{\omega(S(a_1,\bar{a}_j))}{\omega(a_1)}
\,z\,P_{\bar{a}_1 S(\bar{a}_1,a_j)}^<\left(\frac{1}{z}\right)
\Mnu\left(zk_1,k_2;\,..\right)
\nonumber \\*&&\phantom{aa}\!\!
-(-)^{[\sigma(a_1)+\sigma(S(a_1,\bar{a}_j)]}\,\frac{4\pi\as}{k_1\cdot k_j}\,
\frac{\omega(S(a_1,\bar{a}_j))}{\omega(a_1)}
\,z\,Q_{\bar{a}_1 S(\bar{a}_1,a_j)^\star}\left(\frac{1}{z}\right)
\tilde{{\cal M}}^{(n-1)}\left(zk_1,k_2;..,k_j,..\right),
\nonumber \\*&&
\label{crossed}
\eeqn
where $\sigma(g)=0$ and $\sigma(q)=1$ (therefore, $(-)^\sigma$ takes
into account the crossing of a fermionic line), and the overall minus
sign is due to the $\bar{k}_1\to k_1$ substitution.
We can now exploit the crossing symmetry property of the
Altarelli-Parisi kernels (for $z<1$)
\beq
P_{ba}^<(z)=-(-)^{[\sigma(a)+\sigma(b)]}\,\frac{\omega(b)}{\omega(a)}
\,z\,P_{\bar{a}\bar{b}}^<\left(\frac{1}{z}\right)\,.
\eeq
For consistency with this equation, we are thus led to {\it define}
(notice the $^\star$ symbol)
\beq
Q_{b^\star a}(z)=-(-)^{[\sigma(a)+\sigma(b)]}\,\frac{\omega(b)}{\omega(a)}
\,z\,Q_{\bar{a}\bar{b}^\star}\left(\frac{1}{z}\right)\,.
\label{Qcrdef}
\eeq
Eq.~(\ref{crossed}) becomes therefore
\beqn
\Mn\left(k_1,k_2;\,..,k_j,..\right)&\LCu&
\frac{4\pi\as}{k_1\cdot k_j}
\,P_{S(a_1,\bar{a}_j)a_1}^<(z)\Mnu\left(zk_1,k_2;\,..\right)
\nonumber \\*
&+&\frac{4\pi\as}{k_1\cdot k_j}\,Q_{S(a_1,\bar{a}_j)^\star a_1}(z)
\tilde{{\cal M}}^{(n-1)}\left(zk_1,k_2;\,..,k_j,..\right),
\phantom{aa}
\label{initial}
\eeqn
where, from the definition in eq.~(\ref{Qcrdef}), and using
eqs.~(\ref{Q1})-(\ref{Q4}),
\beqn
Q_{g^\star g}(z)&=&-4\CA\,\frac{1-z}{z}\,,
\label{Qstar1}
\\
Q_{q^\star g}(z)&=&0\,,
\\
Q_{g^\star q}(z)&=&-4\CF\,\frac{1-z}{z}\,,
\\
Q_{q^\star q}(z)&=&0\,.
\label{Qstar4}
\eeqn
In spite of the fact that the $\Delta$ term has no deep physical
meaning, still the crossing symmetry property has to hold true,
and the relation between $Q_{ab^\star}$ and $Q_{b^\star a}$
should not be regarded as purely incidental, that is, due
to the definition (\ref{Qcrdef}). In fact, directly performing the
calculation for the initial state collinear splitting, and
defining $Q_{b^\star a}$ through eq.~(\ref{initial}), we
obtain again eqs.~(\ref{Qstar1})-(\ref{Qstar4}).

We finally list the results for $\tilde{{\cal M}}^{(5)}$ for processes
in which only massless partons are involved. This kind of processes
are relevant for three-jet production. The corresponding Born
amplitudes squared, ${\cal M}^{(5)}$, are well known from the literature
(see for example ref.~[\ref{Berendsetal}]).

We start with the purely gluonic processes, denoting the momenta of the
gluons which do not split by $k,l,m,n$, and with $P$ the momentum of
the gluon which eventually splits. We get
\beqn
\tilde{\cal M}^{(5g)}&=&
\frac{1}{2 k_1 \cdot k_2}
\frac{4g_{\sss S}^6 N_c^3 (N_c^2-1)}{\omega(a_1) \omega(a_2)}
\nonumber \\&\times &
{\rm Re}\  \Bigg\{
 \frac{\A i j}{\B i j}
 \left(  \A k l^4  \A m n^4 +
         \A k m^4  \A l n^4 +
         \A k n^4  \A l m^4 \right)
\nonumber  \\&&
   \qquad  \times  \sum_{\sigma^{*}(k,l,m,n)}
    \Big(\A k l \A l m  \A m n \A n P \A P k \Big)^{-2} \Bigg\}.
\eeqn
Here $\sigma^{*}(k,l,m,n)$ denotes all the permutations of the
elements $k,l,m,n$ which are inequivalent under reflection
$(\{klmn\}\to\{nmlk\})$. For the processes with two quarks and three gluons
we denote by $m$ and $n$ the momenta of the gluons which do not participate
in the splitting, and by $q$ and $\bar{q}$ the momenta of the
quark-antiquark pair. We have
\beqn
\tilde{\cal M}^{(2q3g)}&=&
-\frac{1}{2 k_1 \cdot k_2}\frac{4g_{\sss S}^6 }{\omega(a_1) \omega(a_2)}
\nonumber \\*&\times&
{\rm Re} \ \Bigg\{ \frac{\A i j}{\B i j}
 \frac{\A q m \A \qb m \A \qb n \A q n}{\A \qb q^2}
 \left(  \A \qb m^2  \A q n^2 +
         \A q m^2  \A \qb n^2  \right)
\nonumber  \\*&&
   \times \Bigg[ \frac{(N_c^2-1)^3}{2N_c^2}
          \sum_{\sigma(m,n,P)}
               \Big(\A \qb m \A m n \A n P \A P q \Big)^{-2}
\nonumber  \\*&&\phantom{\times}
  +\frac{(N_c^2-1)^2}{N_c^2}
     \sum_{\sigma^*(m,n,P)} \left(
      \frac{1}{\A \qb m \A m n \A n P \A P \qb \A q n^2 \A m P^2} +
           (q \leftrightarrow \qb)\right)
\nonumber  \\*&&\phantom{\times}
  +(N_c^2-1)\sum_{\sigma^*(m,n,P)}
        \frac{1}{\A \qb m \A m q \A q P \A P \qb \A m n^2 \A P n^2 }
       \Bigg] \Bigg\}.
\eeqn
Finally, for processes with four quarks and one gluon, we have to
distinguish between the case in which there are two quark-antiquark
pairs of different flavour, and the case in which the flavour of
the two pairs is the same. In both cases, we denote the
momenta of the quarks by $q$, $\bar{q}$, $Q$ and $\bar{Q}$. The
momentum of the splitting gluon is again $P$.
For quark-antiquark pairs of different flavour we get
\beqn
\tilde{\cal M}^{(4q1g)}_{\rm \sss DF}&=&
\frac{1}{2 k_1 \cdot k_2} \frac{2g_{\sss S}^6}{\omega(a_1) \omega(a_2)}
\ {\rm Re} \Bigg\{ \frac{\A i j}{\B i j}
 \frac{\A \qb \Qb^2 \A q Q^2 + \A \qb Q^2 \A q \Qb^2}
                      {\A \qb q^2 \A \Qb Q^2}
\nonumber  \\*&&
  \times  \Bigg[ \left(N_c^3 - N_c \right) \bigg(
      \frac{\A \qb Q^2}{\A Q P^2 \A \qb P^2} +
      \frac{\A q \Qb^2}{\A q P^2 \A \Qb P^2} \bigg)
\nonumber  \\*&&
  \quad +\ \frac{N_c^2 - 1}{N_c} \bigg(
      \frac{\A \qb q^2}{\A \qb P^2 \A q P^2} -
      \frac{2\ \A \qb q \A \qb Q}{\A \qb P^2 \A q P \A Q P} -
      \frac{2\ \A q \qb \A q \Qb}{\A q P^2 \A \qb P \A \Qb P}
\nonumber  \\*&&
   \qquad \qquad \quad  + \ \frac{\A \Qb Q^2}{\A \Qb P^2 \A Q P^2} -
      \frac{2\ \A q \Qb \A Q \Qb}{\A \Qb P^2 \A q P \A Q P} -
      \frac{2\ \A \qb Q \A \Qb Q}{\A Q P^2 \A \qb P \A \Qb P} \bigg)
\Bigg] \Bigg\}.
\nonumber \\*&&
\eeqn
For quark-antiquark pairs of equal flavour we have
\beqn
\tilde{\cal M}^{(4q1g)}_{{\rm \sss EF}}&=&
\tilde{\cal M}^{(4q1g)}_{{\rm \sss DF}}
+\tilde{\cal M}^{(4q1g)}_{{\rm \sss DF}}(q\leftrightarrow Q)
\nonumber \\*&-&
\frac{1}{2 k_1 \cdot k_2}
\frac{4g_{\sss S}^6\left( N_c^2 - 1 \right)}{\omega(a_1) \omega(a_2)}
\ {\rm Re} \Bigg\{ \frac{\A i j}{\B i j} \ \frac{\A \qb \Qb^2 \A q Q^2}
       {\A \qb q \A \qb Q \A q \Qb \A \Qb Q}
\nonumber \\*&&\times
\bigg[
 \frac{\A \qb q^2}{\A \qb P^2 \A q P^2} +
 \frac{\A \qb Q^2}{\A \qb P^2 \A Q P^2} +
 \frac{\A \Qb q^2}{\A \Qb P^2 \A q P^2} +
 \frac{\A \Qb Q^2}{\A \Qb P^2 \A Q P^2}
\nonumber \\*&&
  \;\;-\frac{N_c^2 + 1}{N_c^2} \
 \Bigg( \frac{\A \qb q \A \qb Q}{\A \qb P^2 \A q P \A Q P} +
        \frac{\A \Qb q \A \Qb Q}{\A \Qb P^2 \A q P \A Q P}
\nonumber \\*&&
   \qquad  \qquad \quad \;+
       \frac{\A q \qb \A q \Qb}{\A q P^2 \A \qb P \A \Qb P} +
       \frac{\A Q \qb \A Q \Qb}{\A Q P^2 \A \qb P \A \Qb P} \bigg)
   \bigg] \Bigg\}.
\eeqn
\begin{reflist}
\item \label{ppreview}
For recent reviews see: \\
 J.~E.~Huth and M.~L.~Mangano, \ar{42}{92}{251};\\
 R.~K.~Ellis and W.~J.~Stirling, Fermilab-Conf-90/164-T (1990) (unpublished).
\item \label{onejettheory}
 S.~D.~Ellis, Z.~Kunszt and D.~E.~Soper, \prl{62}{88}{726};\\
 \prl{64}{90}{2121};\\
 F.~Aversa, M.~Greco, P.~Chiappetta and J.~P.~Guillet,
 \pl{B210}{88}{225};\\
 \pl{B211}{88}{465};
 \np{B327}{89}{105};
 \zp{C46}{90}{253};\\
 \prl{65}{90}{401}.
\item \label{photojettheory}
 D. B\"odeker, G. Kramer, S.~G.~Salesh, \zp{C63}{94}{471};\\
 J.~R.~Forshaw and R.~G.~Roberts, \pl{B348}{95}{665};\\
 L.~E.~Gordon and J.~K.~Storrow, \pl{B319}{93}{539};\\
 M.~Klasen and  G.~Kramer, preprint DESY-95-159, hep-ph/9508337.
\item \label{twojettheory}
 S.~D.~Ellis, Z.~Kunszt and D.~E.~Soper, \prl{69}{92}{1496};\\
 W.~T.~Giele, E.~W.~N.~Glover and D.~A.~Kosower, \prl{73}{94}{2019};\\
 S.~D.~Ellis and  D.~E.~Soper, \prl{74}{95}{5182}.
\item \label{onejet5401800}
 CDF Collaboration, F.~Abe {\it et al.}, \pr{D45}{92}{1448}.
\item \label{onejetexp}
 CDF Collaboration, F.~Abe {\it et al.}, \prl{68}{92}{1104};\\
 \prl{70}{93}{1376}.
\item \label{zeussjet}
 ZEUS Collaboration, M.~Derrick {\it et al.}, \pl{B342}{95}{417};\\
 \pl{B348}{95}{665}.
\item \label{H1jet}
 H1 Collaboration, I.~Abt {\it et al.}, \pl{B314}{93}{436}.
\item \label{twojetexp}
 CDF Collaboration, F.~Abe {\it et al.}, \np{B269}{86}{445};\\
 \pr{D41}{90}{1722}; \prl{68}{92}{1104};\\
 D0 Collaboration, F.~Nang {\it et al.},
 Fermilab Report \\No.~FERMILAB-Conf-94/323-E (1994).
\item \label{hustonetal}
 J.~Huston {\it et al.}, preprint CTEQ-512 (November, 1995).
\item \label{multijetexp}
 UA1 Collaboration, G. Arnison {\it et al.}, \pl{B158}{85}{494};\\
 UA2 Collaboration, J.~A.~Appel {\it et al.}, \zp{C30}{86}{341};\\
 CDF Collaboration, F. Abe {\it et al.}, \pr{D45}{92}{1448};\\
 \prl{75}{95}{608};\\
 D0 Collaboration, S.~Abachi {\it et al.}, hep-ph/9508337;.
\item \label{BDK5g}
 Z.~Bern, L.~Dixon and D.~A.~Kosower, \prl{70}{93}{2677}.
\item \label{KST4q1g}
 Z.~Kunszt, A.~Signer and Z.~Tr\'ocs\'anyi, \pl{B336}{94}{529}.
\item \label{BDK2q3g}
 Z.~Bern, L.~Dixon and D.~A.~Kosower, \np{B437}{95}{259}.
\item \label{ASthesis}
 A.~Signer, PhD Thesis, ETH-Zurich, 1995.
\item \label{sixparton}
 J.~Gunion and Z.~Kunszt, \pl{159B}{85}{167};
 \pl{161B}{85}{333};
 \pl{176B}{85}{163};\\
 Z.~Kunszt, \np{B271}{86}{333};\\
 S.~J.~Parke and T.~R.~Taylor, \np{B269}{86}{410};\\
 M.~L.~Mangano, S.~J.~Parke and Z.~Xu, \np{B298}{88}{653};\\
 J.~Gunion and J.~Kalinowski, \pr{D34}{86}{2119};\\
 F.~A.~Berends and W.~T.~Giele, \np{B306}{88}{759}.
\item \label{Kuijf}
 J.~G.~M.~Kuijf, PhD thesis, Leiden 1991.
\item \label{ManganoParkepr}
 M.~L.~Mangano and S.~J.~Parke, \prep{200}{91}{301}.
\item \label{kssub}
 Z.~Kunszt and D.~E.~Soper, \pr{D46}{92}{192}.
\item \label{EKSgluon}
 S.~D.~Ellis, Z.~Kunszt and D.~E.~Soper, \pr{D40}{89}{2188}.
\item \label{GGcone}
 W.~T.~Giele and E.~W.~N.~Glover, \pr{D46}{92}{1980};\\
 W.~T.~Giele, E.~W.~N.~Glover and D.~A.~Kosower, \np{B403}{93}{633}.
\item \label{HVQetal}
 B.~Mele, P.~Nason and G.~Ridolfi, \np{B357}{91}{409};\\
 M.~Mangano, P.~Nason and G.~Ridolfi, \np{B373}{92}{295};\\
 S.~Frixione, P.~Nason and G.~Ridolfi, \np{B383}{92}{3};\\
 S.~Frixione, \np{B410}{93}{280};\\
 S.~Frixione, M.~Mangano, P.~Nason and G.~Ridolfi , \np{B412}{94}{225}.
\item \label{CSS}
 J.~C.~Collins, D.~E.~Soper and G.~Sterman, in {\it Perturbative
 Quantum Chromodinamics}, 1989, ed. Mueller, World Scientific,
 Singapore, and references therein.
\item \label{ESalg}
 S.~D.~Ellis and D.~E.~Soper, \pr{D48}{93}{3160};\\
 see also: S.~Catani, Yu.~Dokshitser, M.~A.~Seymour and B.~R.~Webber,\\
 \np{B406}{93}{187}.
\item \label{D0report}
 D0 Collaboration, proceedings of the 8$^{th}$ meeting of the American
 Physical Society, Albuquerque, NM, August 2$^{nd}$-~6$^{th}$, 1994.
\item \label{snowmass}
 F.~Aversa {\it et al.}, Proceedings of the Summer Study on
 High Energy Physics, Research Directions for the Decade,
 Snowmass, CO, Jun 25 - Jul 13, 1990.
\item \label{KSTsing}
 Z.~Kunszt, A.~Signer and Z.~Tr\'ocs\'anyi, \np{B420}{94}{550}.
\item \label{ESscale}
 R.~K.~Ellis and J.~Sexton, \np{B269}{86}{445}.
\item \label{KST2to2}
 Z.~Kunszt, A.~Signer and Z.~Tr\'ocs\'anyi, \np{B411}{94}{397}.
\item \label{Berendsetal}
 F.~A.~Berends {\it et al.}, \pl{B103}{81}{124}.

\end{reflist}

\end{document}